\documentclass[prd,aps,preprint,amsmath,nofootinbib,amssymb,eqsecnum,showkeys,tightenlines]{revtex4-1}
\pdfoutput=1
\usepackage{tabularx}
\usepackage{slashed}
\usepackage{epsfig,latexsym,cancel,amssymb,amsmath,verbatim,mathrsfs}
\usepackage{color}
\usepackage{graphicx}
\usepackage{bm}
\usepackage{tabularx}
\usepackage{amsmath}
\usepackage{amssymb}
\usepackage{amsfonts}
\usepackage{cases}
\usepackage{cancel}
\usepackage{hyperref}
\usepackage{ulem}
\usepackage{here}
\usepackage{color}
\usepackage{graphicx}
\usepackage{diagbox}
\usepackage{multirow}
\usepackage{slashed}
\usepackage{simpler-wick}
\usepackage{subfigure}
\usepackage{svg}
\usepackage[marginal]{footmisc}
\usepackage{feynmf}

\def\L{\mathcal{L}}

\def\vec{\mathbf}

\def\L{\left(}
\def\R{\right)}
\def\wt{\widetilde}

\def\f{\frac}
\newcommand{\be}{\begin{equation}}
\newcommand{\ee}{\end{equation}}
\newcommand{\bea}{\begin{eqnarray}}
\newcommand{\eea}{\end{eqnarray}}
\newcommand{\ba}{\begin{array}}
\newcommand{\ea}{\end{array}}

\def\L{\left(}
\def\R{\right)}
\def\wt{\widetilde}

\def\f{\frac}

\long\def\symbolfootnote[#1]#2{\begingroup%
\def\thefootnote{\fnsymbol{footnote}}\footnote[#1]{#2}\endgroup}

\newcommand{\diracslash}[1]{#1\!\!\!/}

\DeclareMathOperator{\Tr}{Tr} \DeclareMathOperator{\diag}{diag}

\newcommand{\beq}{\begin{equation}}
\newcommand{\eeq}{\end{equation}}

\begin{document}

\title{Dark Chiral Phase Transition Driven by Chemical Potential and its Gravitational Wave Test}

\author{Zhaofeng Kang}
\email[E-mail: ]{zhaofengkang@gmail.com}
\affiliation{School of physics, Huazhong University of Science and Technology, Wuhan 430074, China}

\author{Jiang Zhu}
\email{jackpotzhujiang@gmail.com}
\affiliation{Tsung-Dao Lee Institute and  School of Physics and Astronomy, Shanghai Jiao Tong University,
800 Lisuo Road, Shanghai, 200240 China}
\affiliation{Shanghai Key Laboratory for Particle Physics and Cosmology, 
Key Laboratory for Particle Astrophysics and Cosmology (MOE), 
Shanghai Jiao Tong University, Shanghai 200240, China}

\date{\today}

\begin{abstract}

In this article, for the first time, we explore the scenario that the dark-QCD sector has a large chemical potential $\mu$ (on the order of magnitude of temperature) of dark quarks. It leads to a complex-valued Polyakov loop and tilts the partial confinement effect, driving the dark-QCD phase transition to a first-order one in the early universe. We present a toy model via the Affleck-Dine mechanism that could generate degenerate dark quarks. Our study, in the framework of PNJL, focuses on the dynamical impacts of a large chemical potential on the chiral phase transition without turning on the KMT instanton term. We plot the phase diagram of the dark-QCD in the chiral limit. The resulting first-order phase transition actually refers to a chiral phase transition, with the transition to the confinement vacuum being a cross-over. Following the phase diagram, we find that increasing $\mu$ can considerably prolong the duration of the phase transition and also the release of latent heat, which together make the cosmic dark-QCD phase transition at the critical temperature above 1 GeV and below 100 GeV produce gravitational wave signal in the intermediate frequency band, which is well probable in space detectors such as BBO.

\end{abstract}


\maketitle

\section{Introduction}

Beyond standard models (BSM) are usually based on some new symmetries; therefore, they may experience a first-order phase transition (FOPT) in the early universe, which is supposed to release gravitational waves (GW), leaving a component of the stochastic GW background (SWGB) in the universe today due to its decoupling very early. The discovery of such kind of signal would be a very clean smoking gun for new physics since there are no backgrounds from the standard model.

Among various BSMs with FOPT, the dark-QCD sector is of particular interest since the order of the QCD-like phase transition is controlled by the intrinsic non-Abelian dynamics rather than by the concrete model parameters as in the electroweak-like models. The confinement phase transition in the pure Yang-Mills system, such as $SU(N)$ with $N>2$, is known to be first-order and receives a lot of attention~\cite{Halverson:2020xpg, Huang:2020crf, Kang:2021epo, Morgante:2022zvc, Fujikura:2023lkn, Li:2023xto, He:2023ado}. However, adding dynamic quarks may change the order of the phase transition, and it is well known that the QCD phase transition in the standard model with a small chemical potential is a crossover~\cite{Aoki:2006we, Bhattacharya:2014ara, Ding:2015ona}. But this is not the case in dark-QCD where the chiral phase transition, studied in the Nambu Jona-Lasinio (NJL)  model~\cite{Nambu:1961tp, Nambu:1961fr}, may become first-order due to a large KMT determinant term~\cite{Helmboldt:2019pan, Reichert:2021cvs, Pasechnik:2023hwv, Chen:2024jet, Shao:2024dxt, Shao:2024ygm}. However, whether it can be that large is questionable, as it is generated by the instanton effect~\cite{Klevansky:1992qe}. Anyway, it is not the case in real QCD where the experiment data implies that the dimensionless coupling of KMT term is usually not much larger than the dimensionless four-quarks coupling ~\cite{Kohyama:2016fif}.

Alternatively, if there is a large chemical potential, the chiral phase transition will also become a first-order transition~\cite{Asakawa:1989bq, Stephanov:1998dy, Allton:2002zi, Stephanov:2004wx, Ratti:2005jh, Ratti:2006gh, Roessner:2006xn, Zhang:2006gu, Ghosh:2006qh}, but again it cannot be the case in the visible world. Our world today is characterized by high entropy $\eta\equiv n_B/s\sim 10^{-10}$, which serves as the initial input for successful predictions of BBN~\cite{ParticleDataGroup:2020ssz}. Through the argument of entropy conservation, we conclude that the maximum baryon asymmetry today stems from a tiny quark asymmetry $(n_q-n_{\bar q})/n_q\sim n_B/n_\gamma\sim \eta$ in the early universe at temperature above 38 MeV. This explains why we do not need to consider the chemical potential for the light quarks when studying the quark-hadron phase transition in standard thermal cosmology. The large chemical potential is relevant only in nuclear matter, such as neutron stars. However, there may be a large lepton asymmetry $Y_L=\sum_{a=e,\mu,\tau}(n_a+n_{\nu_a})/s$ in our universe, which would also affect the chiral phase transition by the chemical equilibrium of lepton and quarks, causing a large QCD chemical potential (also called the ``Sphaleron freeze-in" effect), and then turning the chiral phase transition into the first-order one~\cite{Gao:2021nwz,Gao:2023djs}. The lepton asymmetry is less constrained by the experimental data.  Although the lepton asymmetry can be much larger than the quark asymmetry, it still suffers the constraint of the BBN, which gives the upper bound $|Y_L|<1.2\times 10^{-2}$~\cite{Serpico:2005bc,Popa:2008tb,Simha:2008mt,Oldengott:2017tzj,Pitrou:2018cgg,Gelmini:2020ekg}. Consequently, the resulting GW signal is quite weak and unable to be detected in future space-based GW detecters~\cite{Zheng:2024tib}.


Let us again turn our attention to the dark-QCD world, where the dark-quark asymmetry is free of any constraint. Therefore, it is of importance to investigate dark chiral phase transition  driven by a large chemical potential. It may have practical meaning related to the visible asymmetry or simply a toy model for the pioneering study of the chemical potential effect in the dark chiral phase transition and the associated GW signal. Our theoretical tool is the conventional Polyakov-loop-extended NJL (PNJL)~\cite{Fukushima:2003fw, Fukushima:2008wg, Ratti:2005jh, Ratti:2006gh, Roessner:2006xn, Zhang:2006gu, Ghosh:2006qh} model with a large chemical potential. We plot the dark-QCD phase diagram in the chiral limit and without turning on the KMT term, determining the critical point for several values of the strength of the four-quark term, $G_S$. Our study shows that the dark-QCD phase transition essentially is reduced to the dark chiral phase transition. Moreover, we analyze the behavior of the phase diagram and try to understand the basic reasons, to find that it is due to the complicated interplay among the NJL model, the confinement effect, and the chemical potential, lacking a simple understanding. However, we still gain insight into the prominent role of chemical potential in the dark chiral phase transition driven in the early universe. Our analysis reveals that the critical temperature $T_c$ is dynamically tuned to finely balance the effective potentials of zero and finite temperature, resulting in a very short phase transition duration time scale in the QCD-like phase transition. However, a large chemical potential is capable of relaxing the tuning and then leads to a much longer phase transition duration, which distinguishes our dark-QCD phase transition from others, showing a more promising detection prospect via the GW signal.   


This paper is organized as follows. In the section.~\ref{LAG}, we will show how to generate the large dark quark asymmetry by the Afflect-Dine mechanism. The next section will present a detailed analysis of the dark chiral phase transition through the PNJL model. Then, we will discuss this phase transition in the early universe and predict the corresponding observable GWs in the section.~\ref{PTGW}. Finally, conclusions, discussions, and the appendix are cast in the remaining two sections. 

\section{Large dark quark asymmetry genesis}\label{LAG}

Let us consider a dark-QCD with $SU(N)$ gauge group and $N_f$ flavors dynamical quarks in the fundamental representation, 
\begin{equation}\label{}
    \mathcal{L}_{DQ}=-\frac{1}{4}F_{\mu\nu}^aF^{\mu\nu a}+\bar{\psi}^i_\alpha\left(i\slashed \partial \delta_{ij}+g\slashed A^a T^a_{ij}-m_\alpha \delta_{ij}\right)\psi^j_\alpha,
\end{equation}
where $i,j=1,2,...N$ denote for the color index and $\alpha$ the flavor index. Classically, it respects the chiral flavor symmetry 
\begin{align}
    SU(N_f)_V\times SU(N_f)_A\times U(1)_V\times U(1)_A,
\end{align}
with $U(1)_A$ explicitly broken by quantum anomaly. Low energy QCD is characterized by the spontaneous breaking of $SU(N_f)_A$ via the strong QCD dynamics on the low energy scale, and the corresponding phase transition is the focus of this article.

The remaining $U(1)_V$ is the conserved baryon number $B$, associated with the quark asymmetry $n_\psi-n_{\bar \psi}$. The $B$ charge of $\psi$ is normalized to be one unit. In this section, we utilized the Afflect-Dine (AD) mechanism~\cite{Affleck:1984fy} to generate a large asymmetry in the early universe. The key ingredient of this mechanism is the presence of a complex field $\phi$ that carries $B$ charge $b$ and, moreover, participates in interactions explicitly breaking $U(1)_B$. There are many options for $\phi$, depending on the model buildings, and here we choose the simplest $\phi_A$ which lies in the antisymmetric rank two tensor representation of $SU(N)$, with $b=-2$; the dimension is $d_A=(N-1)N/2$. This serves as a unifying option for any $N$. It is just a toy model to demonstrate the existence of model, so we bare its drawbacks. Then, the interesting Lagrangian reads 
\begin{equation}\label{}
\begin{split}
    \mathcal{L}_{AD}=&|D_\mu\phi_A|^2-m^2|\phi_A|^2+\lambda |\phi_A|^4+\left(\epsilon |\phi_A|^2{\rm Tr}(\phi_A^2)+\delta {\rm Tr}(\phi_A^4)+c.c.\right) \\
    &
    +\left(y^L_{nm} \phi^{ij}_A \psi^T_{nLi} C\psi_{mLj}+y^R_{nm} \phi^{ij}_A \psi^T_{nRi} C\psi_{mRj}+c.c.\right),
\end{split}
\end{equation}
where the Yukawa coupling matrices satisfy $y^{L/R}_{nm}=-y^{L/R}_{mn}$ since the couplings are antisymmetric after exchanging the flavor index. The terms in the brackets explicitly break $B$, introducing two tiny complex parameters $\epsilon$ and $\delta$, which together admit the necessary CP violation. But if CP is violated by the initial condition, a single term actually works. 

Let us briefly review the AD mechanism. In general, for a classical Lagrangian conserving $U(1)_B$, the charge (density) is the temporal component of the Noether current, $n_B=ib (\dot\phi^*\phi-\phi^*\dot \phi)$. 
Writing $\phi=\phi_R+i\phi_I$, one finds that $n_B\propto \phi_R^2 d\left({\phi_I}/{\phi_R}\right)/dt$, which is nonvanishing only for the non-synchronous evolution of real and imaginary parts, namely the angular motion. For a homogeneous scalar field in the FRW background, its motion follows the equation 
\begin{equation}\label{}
\begin{split}
    \ddot \phi +3H \dot \phi +\partial V/\partial\phi =0,
\end{split}
\end{equation}
from which one can derive the evolution of the charge density stored in the field $\phi$:  
\begin{equation}\label{}
\begin{split}
    \dot n_\phi+3H  n_\phi=a^{-3}d (a^3 n_\phi)/dt=\phi\partial V/\partial\phi -\phi ^*\partial V/\partial\phi^* =2{\rm Im}(\phi \partial V/\partial\phi ).
\end{split}
\end{equation}
The source term does not receive a contribution from the term that conserves the charge, e.g., $\lambda |\phi|^4$. The imaginary part of $\phi\partial V/\partial\phi$, namely CP violation, can be attributed to either complex couplings or the field itself. The former will lead to the latter. The comoving charge density at time $t$ is obtained by integrating over the source term,
\begin{equation}\label{}
\begin{split}
    a^3n_\phi(t)=2\int_{t_i}^ta^3(t'){\rm Im}(\phi\partial V/\partial\phi)dt',
\end{split}
\end{equation}
with $t_i\ll 1/m_\phi$ the initial time when the AD field is at rest, $\dot \phi(t_i)=0$. In our model, 
 \begin{equation}\label{}
\begin{split}
    \partial V/\partial\phi^{ij}_A&=\epsilon\left[(\phi^{ij}_A)^*{\rm Tr}(\phi_A^2)-2|\phi^{ij}_A|^2\phi^{ij}_A\right]+4\delta \phi^{jk}_A\phi^{kl}_A\phi^{li}_A,\\
    \partial V/\partial(\phi^{ij}_A)^*&=\epsilon^*\left[(\phi^{ij}_A){\rm Tr}(\phi_A^2)^*-2|\phi^{ij}_A|^2(\phi^{ij}_A)^*\right]+4\delta^* (\phi_A^{jk}\phi_A^{kl}\phi^{li}_A)^*.
\end{split}
\end{equation}
We take the maximal initial CP violation with $\phi^{ij}_{A,I}(t_i)=\phi^{ij}_{A,R}(t_i)=\phi_0 $. Then we can take $\delta\to0$ and thus each component of $\phi_A$ behaves independently in baryon genesis, and the final result should be understood as a summation of all the $d_A$ components. So, for simplicity, hereafter we will drop the color index and the subscript in $\phi_A^{ij}$.

In the free-field limit, the AD field is almost frozen at their initial position and thus there is little charge genesis. However, the scaling factor increases as $a(t)=C_p t^p$ while the Hubble parameter decreases as $H(t)=p/t$, with $p=1/2$ and $2/3$ in the radiation dominant era (RD) and matter dominant era (MD), respectively. When $H(t)=p/t\sim m_\phi$, the AD field $\phi^{ij}_{\alpha} (\alpha=I,R)$ gains considerable velocity and starts to oscillate with decreasing amplitude. Concretely, the scaling behavior of the oscillating field is given by
\begin{equation}\label{}
\begin{split}
    \phi_\alpha(z)\sim \frac{\phi_{\alpha,0}}{z^{3/4}}\sin z~({\rm RD}), \quad
    \frac{\phi_{\alpha,0}}{z}\sin z~({\rm MD}),
\end{split}
\end{equation}
with $z\equiv m_\phi t\gtrsim 1$.  The field amplitude rapidly decays and eventually settles down at the origin, the minimal of the potential. The source term soon effectively turns off and the asymmetry is frozen at $z$ of a few. 

The above solution is sufficient to provide an estimation on the order of magnitude of the frozen asymmetry, 
 \begin{equation}\label{}
\begin{split}
   a^3n_\phi(t\gg 1/m_\phi)\sim \epsilon C_{2/3}^{3}\phi_0^4 m_\phi^{-3}~({\rm MD}),\quad \epsilon C_{1/2}^{3}\phi_0^4 m_\phi^{-5/2}~({\rm RD}).
\end{split}
\end{equation}
A large hierarchy between $\phi_0$ and $m_\phi$ can lead to a large asymmetry. Let us suppose that this event happens during the RD era where $t\approx 0.3g_*^{-1/2}M_{\rm PL}/T^2$, and therefore one has 
\begin{equation}\label{}
\begin{split}
\frac{n_\psi-n_{\bar \psi}}{T^3}=\frac{n_B}{T^3}\sim 6.1g_*^{3/4}  \epsilon \left(\frac{\phi_0}{m_\phi} \right)^{5/2} \left(\frac{\phi_0}{M_{\rm PL}}\right)^{3/2},
\end{split}
\end{equation}
with $g_*\sim 100$ the relativistic degrees of freedom. Relativistic dark quarks in the degenerate fermion limit, where $n_\psi\approx \frac{g_\psi}{6\pi^2}\mu^3$, allows a rough estimation
\begin{equation}\label{}
\begin{split}
\frac{\mu}{T}\sim \left(\epsilon\frac{ 18.3\pi^2g_*^{3/4}}{g_\psi}\right)^{1/3}
 \left(\frac{\phi_0}{m_\phi} \right)^{5/6} \left(\frac{\phi_0}{M_{\rm PL}}\right)^{1/2},
\end{split}
\end{equation}
which can be made large by setting $\phi_0$ and $m_\phi$ consistently with a very small $\epsilon$.

\section{phase transition in the highly asymmetric dark-QCD}\label{PTQCD}

Historically, the QCD phase transition was first studied in two individual sectors for two different phase transitions,  the pure gauge sector for the deconfinement-confinement phase transition and the quark sector for the chiral phase transition. The former has a well defined order parameter, the traced Polyakov Loop (PL)~\cite{McLerran:1981pb},\begin{equation}\label{define:Pol}
    l={\Tr_c}L=\frac{\Tr_c}{N}{\cal P}\exp\left(ig\int_0^\beta A^a_4T^ad\tau\right),
\end{equation}
where $\beta=1/T$, ${\rm Tr}_c$ denotes the trace of the color index, ${\cal P}$ is the path order operator, and $A_4$ is the Euclidean temporal component of background gauge field $A_\mu$ in the finite-temperature field theory. For $SU(3)$ gauge theory, the Landau free energy for $l$ can be constructed through the PL model following the underlying $SU(3)$ center $Z_3$~\footnote{We only discuss the $SU(3)$ gauge group here, for simplicity, because it will allow us to relate the value of some quantities by the QCD data, such as $\Lambda$ and $T_0$. It is possible to consider a higher color number with more free parameters.}. The latter has an order parameter charged under chiral symmetry, the quark condensation $\langle \bar{\psi}_i\psi_j\rangle$, whose Landau free energy can be derived from the NJL model. In principle, the actual QCD phase transition involves both of these two order parameters. But how to build a proper model is unclear until the appearance of the PNJL model~\cite{Fukushima:2003fw, Fukushima:2008wg,Fukushima:2017csk}, which successfully incorporates the effect of dynamical quarks on the behavior of PL, and as well the confinement effect when studying the chiral phase transition.

In the following, we will first calculate the effective potential of the order parameters in this model and then plot the phase diagram in terms of it.

\subsection{Effective potential in the PNJL model with a chemical potential}


The matter content of the dark-QCD mimics that of the real QCD, containing three flavors of dynamical quarks in the chiral limit. Then, the dark-QCD respects the chiral symmetries $SU(3)_V\times SU(3)_A\times U(1)_V\times U(1)_A$. As in the real QCD, we take the NJL Lagrangian to model the low-energy dark-QCD dynamics which are complicated due to the non-perturbative effects in the infrared. The model is fairly simple, about several flavors of Dirac fermions in the chiral limit, 
\begin{equation}\label{NJL}
\begin{split}
    \mathcal{L}_{\rm chiral}=\bar{q}i\slashed\partial q+G_S\sum_{a=1}^8&\left[(\bar{q}\lambda^a q)^2+(\bar{q}i\gamma^5\lambda^a q)^2\right]
    \\
    &+G_D[\det(\bar{q}(1-\gamma^5)q+\bar{q}(1+\gamma^5)q)],
\end{split}
\end{equation}
where $q=(u,d,s)$ and $\lambda^a$ the eight Gelmann matrices for flavor $SU(3)$. It does not contain gluons that have been integrated out, leaving effects partially encoded in the dimension 6 operators measured by $G_S$. They may induce spontaneous breaking of the chiral symmetry. The KMT determinant interaction is induced by the instanton effect due to the anomaly of $U(1)_A$, with a strength $G_D$, which is not important in our discusssion. 


%


 On top of that, a gluonic background may form, known as the PL background, now the psuedo order parameter for confinement in the presence of dynamical quarks. We will include this part in the finite temperature.



\subsubsection{Chiral symmetry breaking at zero temperature}

To determine the dark-QCD vacuum from the NJL model, one needs to first calculate the effective potential for the quark condense $\langle \bar{q}_iq_i\rangle$. To that end, let us rewrite the bilinear quark field as the condensate part plus the fluctuation $\bar{q}q=\langle\bar{q}q\rangle+\hat{\bar{q}}\hat{q}$. As a bosonlization of the NJL model, one introduces bosonic collective fields corresponding to the various $\bar q q$ bound states. In particular, the condensed field for chiral symmetry breaking is defined as $\sigma=4G_S\langle\bar{q}q\rangle$, with dimension 1. Following the Hubbard-Stratonovich (HS) transformation~\cite{Hubbard:1959ub}, the effective Lagrangian Eq.~(\ref{NJL}) can be rewritten as
\begin{equation}\label{sigmaNJL}
    \mathcal{L}^{eff}_{NJL}=\bar{q}_i[i\slashed\partial-M_i(\sigma)]q_i-V_{NJL}^{tree}(\sigma).
\end{equation}
Then, it is straightforward to extract the tree level potential of the $\sigma$ field from \eqref{NJL},
\begin{equation}\label{treep}
\begin{split}
    \mathcal{V}_{NJL}^{tree}(\sigma)&=\frac{3\sigma^2}{8G_S}+\frac{G_D\sigma^3}{128G_S^3}.
\end{split}
\end{equation}
There is also some cross-terms between the condense field $\sigma$ and the fluctuation field, which turn out to be the constituent mass term for the quarks, 
\begin{equation}
    M_i(\sigma) = -\sigma - \frac{G_D \sigma^2}{64 G_S^2},
    \label{eq:mass}
\end{equation}
which is universal to all quark flavors.  

In addition to $\mathcal{V}_{NJL}^{tree}$, the order parameter field also receives the contribution from the vacuum energy $\mathcal{V}_{NJL}^{vac}$. It is computed by integrating over the zero-point energy for each quark flavor $\frac{1}{2}\omega_i(p)=\frac{1}{2}\sqrt{\vec{p}^2+M_i^2(\sigma)}$, counting the four degrees of freedom for each Dirac fermion
\begin{equation}\label{vacengy}
\begin{split}
    \mathcal{V}_{NJL}^{vac}&=-4 N N_f\int^\Lambda\frac{d^3\vec{p}}{(2\pi)^3}\frac{1}{2}\sqrt{\vec{p}^2+M^2_i(\sigma)}
    \\
    &=-2N_f N\frac{\Lambda^4}{16\pi^2}\bigg[(2+\zeta^2)\sqrt{1+\zeta^2}+\frac{\zeta^4}{2}\log\left(\frac{\sqrt{1+\zeta^2}-1}{\sqrt{1+\zeta^2}+1}\right)\bigg],
\end{split}
\end{equation}
where $N=N_f=3$, $\zeta\equiv\sigma/\Lambda$, and $\Lambda$ is the necessary UV cut-off of the NJL effective theory, which is non renormalizable; and any mass scale present in the NJL model should lie below it. Notice that in this paper, we applied the 3-D cut-off scheme. The vacuum energy and the meson masses, which are given in the Appendix~\ref{spectrum}, have different expressions for different regularization schemes. In real QCD, $\Lambda$ can be fixed from the meson spectrum, but it is not available in dark-QCD, and we will turn back to this later. 



In terms of the total zero-temperature effective potential for the order parameter field, $V_{NJL}^{zero}(\sigma)=V_{NJL}^{tree}+V_{NJL}^{vac}$, it can be proven that condense occurs only when $G_S\Lambda^2\geq \frac{3}{2} \pi^2/(NN_f))$~\cite{Fukushima:2017csk}. There is another constraint from the theoretical consistency of effective theory: the constituent quark mass $M(\sigma)$ from chiral condensate should not exceed the cut-off scale $\Lambda$, which imposes the bound $G_S\Lambda^2\lesssim 3$. So, the parameter  $G_S\Lambda^2$ of interest lies in the following region $ \pi^2/6\leq G_S\Lambda^2\leq 3$.


\subsubsection{finite-temperature effective potential taking into account the confinement effect}

To study the phase diagram of the model, we need to correctly include the finite-temperature effects in the case of a large chemical potential. At the same time, the confinement effect enters through the formation of a PL background and is felt by the quarks.

The finite-temperature contribution $\mathcal{V}_{NJL}^{T}$ from the quasiquarks can be calculated by the standard procedure of finite-temperature field theory. However, one must notice that, in the presence of a temporal gauge field background $A_0$, the covariant derivative of the quasi quark is $D_\mu=\partial_\mu+iA_0\delta_{\mu 0}$. This is not inconsistent with the spirit of NJL, which is the leading EFT after integrating out the gluon and ghost, because here $A_0$ is simply the background. Then, the thermal excitations of the quark contribute to the free energy of both order parameters, i.e., $V\sim {\rm Tr}\log[D^2(A_0)+M^2(\sigma)]$, and as a result, the NJL model is extended to the PNJL model~\cite{Fukushima:2003fw, Fukushima:2008wg,Fukushima:2017csk}. Let us split  the trace into the momentum space and color space, 
\begin{equation}\label{NJL:T}
    \mathcal{V}_{NJL}^{T}=-\frac{3T^4}{\pi^2}\int dx x^2{\rm Tr}_c\Bigg(\log\bigg[1+Le^{-\sqrt{x^2+\frac{M(\sigma)^2}{T^2}}+\frac{\mu}{T}}\bigg]+\log\bigg[1+L^\dagger e^{-\sqrt{x^2+\frac{M(\sigma)^2}{T^2}}-\frac{\mu}{T}}\bigg]
    \Bigg),
\end{equation}
with $x^2\equiv \vec p^2/T^2$. The two parts denote the contributions from quark and antiquarks, respectively, and are biased by the large chemical potential $\mu\sim T$.  

Substituting the ${\rm Tr_c}$ into the logarithmic function through the identity ${\rm Tr \log A}=\log\det(A)$, and further using the fact that in the Polyakov gauge, where the Polyakov Loops Eq.~(\ref{define:Pol}) can be written as the diagonal phase matrix of the temporal gauge fields associated with the generators in the Cartan algebra
\begin{equation}\label{BGF}
    L=\frac{1}{3}\exp\L\frac{igA_4}{T}\R=\frac{1}{3}\exp\L\frac{igA_4^aT^a}{T}\R=\frac{1}{3}\exp(\diag(i\theta_1,i\theta_2,i\theta_3)),
\end{equation}
with $\theta_1=\frac{g}{2\sqrt{3}T}(A_4^8+\sqrt{3}A_4^3)$, $\theta_2=\frac{g}{2\sqrt{3}T}(A_4^8-\sqrt{3}A_4^3)$ and $\theta_3=-\frac{g}{\sqrt{3}T}A_4^8$. Then, the traced PL is the sum of the three phases $l=(\sum_{n=1}^3e^{i\theta_n})/3$. 


This treatment enables us to manifest the confinement effect of PL background in \eqref{NJL:T} in an illustrative manner
\begin{equation}
\begin{split}
    \mathcal{V}_{NJL}^{T}=-\frac{3T^4}{\pi^2}\int dx x^2 G(x,\sigma,T,\mu)
\end{split}
\end{equation}
where the function $G(x,\sigma,T,\mu)$ is given by
\begin{equation}
\begin{split}
    G(x,\sigma,T,\mu)&=
    \log\Big[1+e^{-3(\sqrt{x^2+\frac{M(\sigma)^2}{T^2}}+\frac{\mu}{T})}+3l e^{-\sqrt{x^2+\frac{M(\sigma)^2}{T^2}}-\frac{\mu}{T}}+3l^*e^{-2(\sqrt{x^2+\frac{M(\sigma)^2}{T^2}}+\frac{\mu}{T})}\Big]
    \\
    &+\log\Big[1+e^{-3(\sqrt{x^2+\frac{M(\sigma)^2}{T^2}}-\frac{\mu}{T})}+3l^*e^{-\sqrt{x^2+\frac{M(\sigma)^2}{T^2}}+\frac{\mu}{T}}+3le^{-2(\sqrt{x^2+\frac{M(\sigma)^2}{T^2}}-\frac{\mu}{T})}\Big].
\end{split}
\end{equation}
In the confinement phase where $l=0$, only the baryonic-like term survives, indicating that the quasi-quark excitation is suppressed.

In addition to the contribution from quasi quarks, the free energy of the PL also receives a contribution from the pure gluonic sector at finite $T$, which may be incalculable via the ordinary perturbative method near the critical temperature. A conventional way is following the Landau approach, to construct the Landau free energy for $l$ which is invariant under the global center $Z_N\in SU(N)$ with $N=3$ here; by contrast, $\mathcal{V}_{NJL}^{T}(l,l^*)$ does not respect this symmetry. We take the one well studied in \cite{Ratti:2005jh}
\begin{equation}
\begin{split}
    V_{PLM}(l,l^*,&T)=T^4\left[\frac{-b_2(T)}{2}ll^*+\frac{b_3}{6}(l^3+{l^*}^3)+\frac{b_4}{4}(ll^*)^2\right]
    \\
    b_2(T)&=a_0+a_1\bigg(\frac{T_0}{T}\bigg)+a_2\bigg(\frac{T_0}{T}\bigg)^2+a_3\bigg(\frac{T_0}{T}\bigg)^3
\end{split}
\end{equation}
The coefficients $b_{3,4}$ and $a_{0,1,2,3}$ are determined by fitting the lattice data of pure $SU(3)$ thermodynamics, and the corresponding result is present in Table.\ref{table-1}.


The above polynomial model has been generalized to unifiedly study the confinement-deconfinement phase transition and thermodynamics for the general hot $SU(N)$ system~\cite{Pisarski:2001pe, Ratti:2006gh, Huang:2020crf, Kang:2021epo}. In particular, following the approach of temperature-dependent quasi-gluons, a successful Landau free energy can be calculated perturbatively~\cite{Reinosa:2014zta, Reinosa:2014ooa, Reinosa:2015gxn, Kang:2022jbg}. We are not sure whether different options of PL models will yield similar results, and let us leave it to future publications.

\subsubsection{The complete PNJL model}

Therefore, the theoretical tool to study dark-QCD phase transition is the following total effective potential for the order parameters,
\begin{equation}\label{VPNJL}
    V_{PNJL}(l,l^*,\sigma,T,\mu)=V_{NJL}^{zero}(\sigma)+V_{NJL}^T(l,l^*,\sigma,T,\mu)+V_{PLM}(l,l^*,T).
\end{equation}
With it, we are capable of plotting the dark-QCD  phase diagram and studying the corresponding dynamics of the phase transition in the early universe. However, before moving on to these topics, we would like to figure out how many free parameters that could affect them. 



In this model, five parameters are undetermined: $\Lambda, T_0, G_S, G_D$, and the chemical potential $\mu$~\footnote{In general, $G_{S,D}$ are supposed to depend on $T$ and $\mu$, but we are lack of a reliable way to fix the dependence and then the parameters determined from vacuum will be adopted, which leads to the largest theoretical uncertainty.}. Among these parameters, the cutoff scale $\Lambda$ and the critical temperature for the pure Yang-Mills sector $T_0$ should be determined by the confinement effect. Therefore, these two parameters should not be independent. In real QCD, the cutoff scale is around $960 ~{\rm MeV}$ for the small $s$-quark mass case \cite{Kohyama:2016fif} and $T_0\sim 260 {\rm MeV}$ in the pure $SU(3)$ Yang-Mills theory, and thus they imply the ratio $\Lambda/T_0\sim3.5$. In this work, we assumed that this ratio also holds in the dark-QCD sector. $\Lambda, G_S, G_D$ are related to the mass of the meson and baryon and their decay rates. In QCD, one can use the spectrum and another observable to determine these few parameters. However, this is not the situation in the dark sector, and thus they are essentially free parameters; instead, the spectrum is a prediction of the theory. But this is not the focus of this work, and we need to introduce proper terms to eliminate those massless GSB modes.

Fortunately, in the chiral limit, it is argued that $G_D {\rm GeV^5}\sim 1/{N^{N_f}}$ in the large $N$ limit~\cite{Osipov:2004mn}. Since the contribution of $G_D$ to condensation $\sigma$ and quasi-quark mass $M(\sigma)$ is proportional to a large depression factor, $G_D$ only contributes to chiral phase transition when it is much larger than $G_S$. So, we will set it to zero and then only three free parameters remain, 
\begin{align}\label{bound}
 \mu,\quad   T_0, \quad \pi^2/6\leq G_S\Lambda^2\leq 3,
\end{align}
with $\mu$ around $T_0$. However, we have to stress that, when $G_D\Lambda^5\gg G_s\Lambda^2$, the KMT determinant interaction will become critical, and can even make the chiral phase transition become first-order at zero-chemical potential \cite{Helmboldt:2019pan, Reichert:2021cvs}.


\subsection{The dark-QCD phase diagram}\label{PDiagram}

Our task in this subsection is to plot the phase diagram for dark-QCD with a large chemical potential based on the PNJL model, which couples the three order parameters $\Phi=(l,l^*,\sigma)$ via $V^T_{PNJL}$. This diagram is drawn by studying the phase transition behavior for different $T/T_0$ and $\mu/T_0$, taking some fixed $G_S\Lambda^2$. To be more specific, first we should determine the order of the phase transition; secondly, we need to find the critical temperature $T_c$ for different parameters.

\subsubsection{Plotting the phase diagram}

To determine the order of phase transition, we need to calculate the values of the three order fields, which are the solution of the coupled equations, 
\begin{equation}\label{findvac}
    \frac{\partial V_{eff}}{\partial l}=0,\quad\quad \frac{\partial V_{eff}}{\partial l^*}=0,\quad\quad \frac{\partial V_{eff}}{\partial \sigma}=0.
\end{equation}
In principle, $l$ and $l^*$ should be complex conjugate fields to each other. In this case, one can treat the real and imaginary part of field $l$ to study the phase structure. However, this procedure will make the effective potential became complex. To our knowledge, this fact is induced by the fermion sign problem~\cite{Fukushima:2017csk}. So, like in the previous literature~\cite{Zhang:2006gu,Ghosh:2006qh,Ratti:2005jh}, we treat $l$ and $ l^*$ as two independent real fields, and then we will explore the vacuum structure in a three-dimensional field space. Suppose that the above equations admit multiple roots $\Phi_i$ and label the global minimum $\Phi_{min}$ (the $l$ and $ l^*$ components of $\Phi_{min}$ are unequal in the presence of $\mu\neq0$), that is, $V_{eff}(\Phi_{min},T,\mu)={\rm Min}[V_{eff}(\Phi_i,T,\mu)]$. As $V_{eff}(\Phi_i,T,\mu)$ changes with decreasing temperature, $\Phi_{min}(T)$ also evolves, and if it is a continuous function, then the phase transition is a crossover one; otherwise, it is a first-order phase transition (FOPT). 
\begin{figure}[htbp]
\centering 
\includegraphics[width=0.45\textwidth]{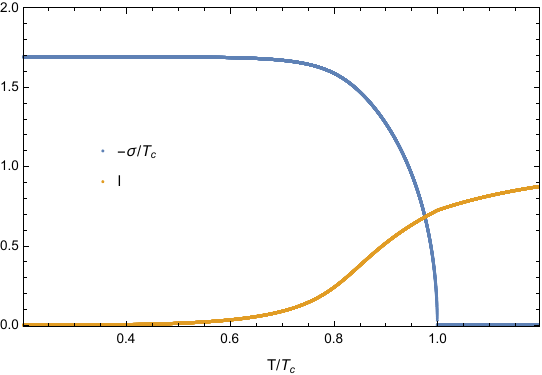}
\includegraphics[width=0.45\textwidth]{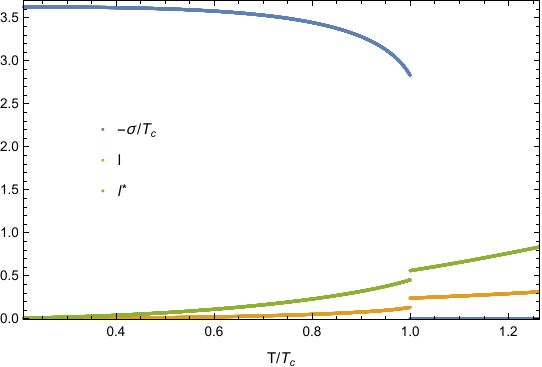}
\caption{The plot for the order parameter as a function of temperature with $G_S=2.2/\Lambda^2$. In the left panel, the chemical potential $\mu=0$ and in this case $l=l^*$; In the right panel, chemical potential $\mu=1.65T_0$ and $l\neq l^*$.} \label{PT:tra}
\end{figure}

We find that just above $T_c$ (determined later), $V_{eff}(\Phi_i,T,\mu)$ has global minima located at the position $\Omega_{0}=(\langle l\rangle_0,\langle l^*\rangle_0,\langle\sigma\rangle=0)$, the deconfinement and non-chiral vacuum; there is a metal stable one located at $\Omega_{1}=(\langle l\rangle_1,\langle l^*\rangle_1,\langle\sigma\rangle_1\neq0)$, the partial confinement and chiral vacuum. A barrier separates them. At $T_c$, the transition from $\Omega_0$ to $\Omega_1$ is made via FOPT.  The process is shown in the right panel of Fig.~\ref{PT:tra}. It can be seen that this FOPT jumps significantly from zero to $\langle \sigma \rangle_1 $ along $\sigma $, however, the jumps along $l$ and $l^*$ are not significant. Below $T_c$, with the continuously decreasing PL and increasing quark condensation, $\Omega_1$ evolves to the complete confinement and chiral vacuum $\Omega_2=(0,0,\langle\sigma\rangle_2)$ as $T$ decreases. In contrast, for the PNJL model with a vanishing chemical potential, we simply have a crossover, as shown in the left panel of Fig.~\ref{PT:tra}, which was well known before~\cite{Fukushima:2017csk}. Note that in either case, the transition from the deconfinement phase to the confinement phase is not a first-order phase transition.

\begin{figure}[htbp]
\centering 
\includegraphics[width=0.45\textwidth]{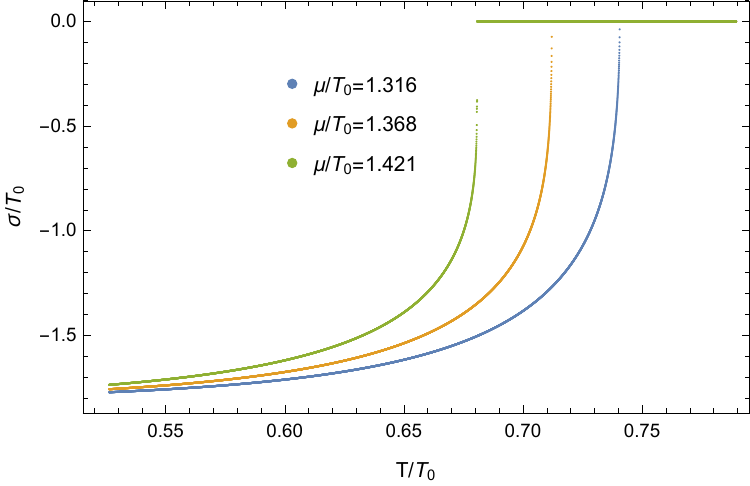}
\includegraphics[width=0.45\textwidth]{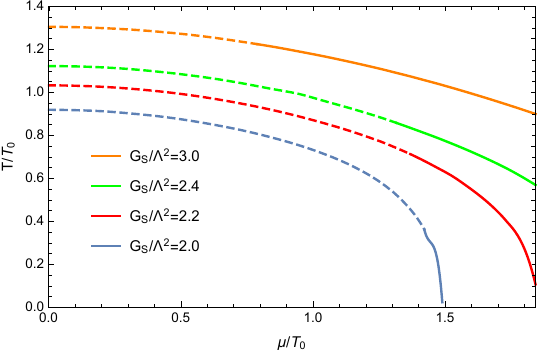} 
\caption{The left panel is the jumping behavior of the order parameter $\sigma$ with the temperature in different chemical potential $\mu$ for $G_S=2.2/\Lambda^2$. The right panel is the phase diagram for different $G_S$, where, following each line, the perpendicular axis is the chemical potential and the vertical axis is the critical temperature for the phase transition.} \label{phsdiag}
\end{figure}

Because the chemical potential is crucial for changing the order of the chiral phase transition, it is necessary to study its role more specifically. We show the behavior of the order parameter $\sigma$ for several values of $\mu$ in the left panel of Fig.~\ref{phsdiag}, choosing $G_S\Lambda^2=2.2$. One can see that when $\mu/T_0=1.421$ and $\mu/T_0=1.368$, the $\sigma$ obviously has discontinuous jumps at $T_c$, then the corresponding chiral phase transition is a FOPT. However, when $\mu/T_0$ decreases to a value $1.316$, $\sigma$ becomes a continuous function of temperature, and then the phase transition changes to a crossover. 

The next step is to solve $T_c$ for different $\mu$ when $G_S$ is fixed. For a crossover transition, as shown in Fig.\ref{phsdiag}, one can start from a lower temperature where $\partial \sigma/\partial T\neq 0$ and slowly increase the temperature and determine $T_c$ by $\partial \sigma/\partial T_c=0$; for the first-order phase transition, it can be calculated by the vacuum-degenerate condition $V_{eff}(\Phi_1, T_c,\mu)=V_{eff}(\Phi_2, T_c,\mu)$ (suppose there are two vacua). Finally, the phase diagram can be plotted by plotting $T_c$ as a function of $\mu$ in the $T-\mu$ plane for some $G_S\Lambda^2$. We choose four benchmark points $G_S\Lambda^2=(2.0,2.2,2.4,3.0)$, labeled as Benchmarks $A$, $B$, $C$, and $D$, details can be found in Appendix.\ref{spectrum}, as depicted in the right panel of Fig.~\ref{PT:tra}, where the dashed line represents a crossover, and the solid line represents FOPT. The dividing point between the two lines is the critical point. The behaviors observed in the $T-\mu$ plane can be attributed to the interplay between the strength of the interactions characterized by $G_S$ and the thermodynamic properties of the system:
\begin{itemize}
    \item For a larger $G_S$, the critical point shifts towards the obviously smaller chemical potential. This results in a more pronounced FOPT. 
    \item   Furthermore, there is a lower bound on $G_S$ to admit a FOPT, and we find that it occurs around $G_S\Lambda^2\approx 1.8$, close to the one indicated by \eqref{bound}. For that small $G_S\Lambda^2$, the viable parameter for FOPT in $T-\mu$ space is too small.
    \item Along each line (with fixed $T_0$), as $\mu$ increases, the critical temperature $T_c$ rapidly decreases and the latent heat release (not shown) also decreases. 
\end{itemize}

It is not easy to get a clear understanding of the underlying reasons for the FOPT chiral phase transition, and it is a consequence of the entanglement between $\sigma$ and PL, further complicated by the chemical potential. To test the role of the confinement effect, we plot the dark-QCD phase diagram working in the NJL with $G_D=0$, to find that the critical point disappears and there is no FOPT.

\subsubsection{Deviate from the main line: chemical potential effect on the electroweak-like phase transition}\label{EW:ex}

We wonder whether the mechanism of dark-QCD FOPT driven by a large chemical potential can be simply translated into an electroweak-like system. To that end, we consider a toy model which consists of a real scalar $\phi$ and a Dirac fermion $\psi$, and they admit the following Lagrangian, 
\begin{equation}
    \mathcal{L}=\frac{1}{2}\partial_\mu\phi\partial^\mu\phi+\frac{\mu^2}{2}\phi^2-\frac{\lambda}{4}\phi^4+i\bar{\psi}\diracslash{\partial}\psi+g\phi\bar{\psi}\psi,
\end{equation}
with $\mu^2>0, \lambda>0$. The field $\psi$ does not have a bare mass term. The pure scalar part respects the $Z_2$ symmetry which acts merely on $\phi$, but the Yukawa term breaks it explicitly. Our discussion of the possible phase transition related to $Z_2$ is in a plasma of particle $\psi$ with a large chemical potential.

The scalar potential at zero temperature does not have a tree-level cubic term, and thus the phase transition is second order. However, the thermal correction from the fermions (not the bosons) may create a barrier and lead to FOPT.  In the case with zero chemical potential, we indeed find that there is a FOPT, if the Yukawa coupling $g$ is large enough~\footnote{This is true up to two-loop level, but some lattice studies indicate that it may be a fake phenomenon caused by the breaking down of the thermal perturbation theory~\cite{Niemi:2024axp}. }. However, it does not persist when we decrease $g$ to see the effect of increasing the chemical potential;  changing the chemical potential merely changes the value of the critical temperature but never turns the high-order phase transition into a FOPT.

\section{The Phase Transition in the early universe}\label{PTGW}

In general, the phase transition in the early universe is governed by the Euclidean action and the bounce equations for the order parameter fields. It is particularly complicated for dark-QCD, as it involves three order parameter fields $\sigma$, $l$, and $l^*$, and thus in principle there are three coupled bounce equations.  Fortunately, a significant simplification can be made in terms of our previous analysis, which tells us that during the chiral phase transition the PL field only experiences a small jump, which has a minor effect. This fact enables us to use the first two equations of Eq.~(\ref{findvac}) to express $l$ and $l^*$ as a function of $\sigma$ and to make the chiral phase transition effectively govern by a single-field bounce equation. In other words, the tunneling problem in the three-dimensional field space is equivalent to a one-dimensional tunneling problem along the gradient path of the effective potential. We will test this simplification in a forthcoming publication showing that the resulting difference by using the complete tunneling path is within one percent~\cite{Hua:2025}.

However, even in this case, the bounce equation for $\sigma$ still differs from the normal scalar field bounce. This is because $\sigma$ is a composite field and does not have kinetic terms at the tree level. This means that the kinetic term for the $\sigma$ field must rise from the loop correction and be proportional to the renormalization factor. In this case, the 3-dimensional Euclidean action is given by
\begin{equation}\label{3daction}
    S_E=4\pi\int dr r^2\left(\frac{1}{2Z_\sigma}\bigg(\frac{d\sigma}{dr}\bigg)^2+V_{PNJL}[\sigma,l(\sigma),l^*(\sigma),T,\mu]\right),
\end{equation}
where $1/Z_\sigma$ is the renormalization factor for $\sigma$. Then, the corresponding bounce equation along with the boundary conditions at the escaping point $r=0$ and the false vacuum point $r\to \infty$, is modified as the following
\begin{align}\label{bounce}
  &  \frac{d^2\sigma}{dr^2}+\frac{2}{r}\frac{d\sigma}{dr}-\frac{1}{2}\frac{Z_\sigma'}{Z_\sigma}\left(\frac{d\sigma}{dr}\right)^2=Z_\sigma V_{PNJL}',\\
   & \frac{d\sigma}{dr}|_{r=0}=0, \quad \sigma|_{r\rightarrow\infty}=0,
\end{align}
where prime denotes the derivative with respect to $\sigma$. As usual, this equation can be solved using the overshooting and undershooting method. However, above all, we should find the expression of the renormalization factor.

\subsection{Renormalization factor for kinetic term with non-zero PL background and $\mu$}

The renormalization factor of the wave function is determined by $\Gamma_{\sigma\sigma}$, the 1-PI 2-point function of the field $\sigma$ as the following,
\begin{equation}
    Z_\sigma^{-1}=-\frac{d\Gamma_{\sigma\sigma}(q_0,\vec{q},\sigma)}{d\vec{q}^2}|_{q_0=0,\vec{q}^2=0},
\end{equation}
which receives contributions from the quasi-quarks. Note that we are considering the thermal tunneling effect at finite $T$, so this renormalization factor should also be calculated in the 3D case~\cite{Lofgren:2021ogg,Hirvonen:2021zej}. In this case, the effect of the background field $A_4$ should also be included, just as when we compute the effective potential. To our knowledge, this effect was first taken into account in \cite{Reichert:2021cvs}. Since this is relatively new, it deserves a calculation with full details that may be helpful for cross-checking.

\subsubsection{Calculation of $\Gamma_{\sigma\sigma}$ at finite $T$}

To calculate $\Gamma_{\sigma\sigma}$, we collect the relevant couplings from the Lagrangian Eq.~(\ref{sigmaNJL}), 
\begin{equation}\label{}
    \mathcal{L}^{eff}_{NJL}=\bar{q}[i\slashed\partial-M(\sigma)+\gamma^0(\mu-iA_4)]q-V_{NJL}^{tree},
\end{equation}
from which one can obtain the following Feynman rules involving the mean field $\sigma$,

\tikzset{every picture/.style={line width=0.75pt}} 

\begin{tikzpicture}[x=0.75pt,y=0.75pt,yscale=-1,xscale=1]

\draw  [dash pattern={on 4.5pt off 4.5pt}]  (22.91,101.58) -- (96.6,101.58) ;
\draw  [dash pattern={on 4.5pt off 4.5pt}]  (269.02,101.58) -- (306.24,101.58) ;
\draw    (306.24,101.58) -- (343.46,135.62) ;
\draw    (306.24,101.58) -- (342.72,74.35) ;
\draw    (481.18,86.83) -- (554.87,122.38) ;
\draw  [dash pattern={on 4.5pt off 4.5pt}]  (481.18,123.14) -- (554.87,87.59) ;

\draw (10.21,89.95) node [anchor=north west][inner sep=0.75pt]    {$\sigma $};
\draw (95.82,89.95) node [anchor=north west][inner sep=0.75pt]    {$\sigma $};
\draw (256.32,90.71) node [anchor=north west][inner sep=0.75pt]    {$\sigma$};
\draw (469.97,113.03) node [anchor=north west][inner sep=0.75pt]    {$\sigma $};
\draw (554.83,75.96) node [anchor=north west][inner sep=0.75pt]    {$\sigma $};
\draw (341.67,125) node [anchor=north west][inner sep=0.75pt]    {$\psi $};
\draw (341.67,62.96) node [anchor=north west][inner sep=0.75pt]    {$\psi $};
\draw (555.32,114.03) node [anchor=north west][inner sep=0.75pt]    {$\psi $};
\draw (468.22,78.47) node [anchor=north west][inner sep=0.75pt]    {$\psi $};
\draw (110.45,88.79) node [anchor=north west][inner sep=0.75pt]    {$=-\frac{3i}{4G_{S}} +\frac{3iG_{D}}{8G_{S}^{3}}\sigma$};
\draw (356.04,91.57) node [anchor=north west][inner sep=0.75pt]    {$=-i+\frac{iG_{D}}{8G_{S}^{2}}\sigma$};
\draw (576.11,93.73) node [anchor=north west][inner sep=0.75pt]    {$=\frac{iG_{D}}{8G_{S}^{2}}$};
\end{tikzpicture}
In the mean-field approximation, we do not need to consider the self-interaction of $\sigma$. Then, the one-loop 2-point function receives the following three contributions:
\tikzset{every picture/.style={line width=0.75pt}} 

\begin{tikzpicture}[x=0.75pt,y=0.75pt,yscale=-1,xscale=1]

\draw  [dash pattern={on 4.5pt off 4.5pt}]  (49,150) -- (199,150) ;
\draw   (302,150) .. controls (302,136.19) and (313.19,125) .. (327,125) .. controls (340.81,125) and (352,136.19) .. (352,150) .. controls (352,163.81) and (340.81,175) .. (327,175) .. controls (313.19,175) and (302,163.81) .. (302,150) -- cycle ;
\draw  [dash pattern={on 4.5pt off 4.5pt}]  (352,150) -- (399,150) ;
\draw  [dash pattern={on 4.5pt off 4.5pt}]  (252,150) -- (302,150) ;
\draw  [dash pattern={on 4.5pt off 4.5pt}]  (451,173) -- (601,173) ;
\draw   (501,148) .. controls (501,134.19) and (512.19,123) .. (526,123) .. controls (539.81,123) and (551,134.19) .. (551,148) .. controls (551,161.81) and (539.81,173) .. (526,173) .. controls (512.19,173) and (501,161.81) .. (501,148) -- cycle ;

\draw (50,128.4) node [anchor=north west][inner sep=0.75pt]    {$\sigma $};
\draw (192,129.4) node [anchor=north west][inner sep=0.75pt]    {$\sigma $};
\draw (252,128.4) node [anchor=north west][inner sep=0.75pt]    {$\sigma $};
\draw (387,129.4) node [anchor=north west][inner sep=0.75pt]    {$\sigma $};
\draw (451,150.4) node [anchor=north west][inner sep=0.75pt]    {$\sigma $};
\draw (588,150.4) node [anchor=north west][inner sep=0.75pt]    {$\sigma $};
\draw (320,103.4) node [anchor=north west][inner sep=0.75pt]    {$\psi $};
\draw (320,153.4) node [anchor=north west][inner sep=0.75pt]    {$\psi $};
\draw (518,100.4) node [anchor=north west][inner sep=0.75pt]    {$\psi $};
\end{tikzpicture}.\\
The straightforward calculation leads to the following expression
\begin{equation}\label{1PI}
    \Gamma_{\sigma\sigma}(q)=-\frac{3}{4G_S}+\frac{3G_D}{8G_S^3}\sigma-9(1-\frac{G_D\sigma}{4G_S^2})^2A(q^2)+9\frac{G_D}{4G_S^2}B(q^2),
\end{equation}
where $A(q^2)$ and $B(q^2)$ are one-loop functions corresponding to the second and third diagrams, respectively
\begin{equation}
\begin{split}
    &A(q^2)=\frac{1}{3}\Tr_c\int\frac{d^4k}{i(2\pi)^4}\frac{\Tr[(\slashed k+\slashed q+M(\sigma))(\slashed k+M(\sigma))]}{[(k+q)^2-M^2(\sigma)][k^2-M^2(\sigma)]},
    \\
    &B(q^2)=\frac{1}{3}\Tr_c\int\frac{d^4k}{i(2\pi)^4}\frac{M(\sigma)}{k^2-M^2(\sigma)}.
\end{split}
\end{equation}

Since we consider the correction to the self-energy of $\sigma$ from both quantum and thermal fluctuations, we should compute the renormalization factor using finite-temperature field theory, where $k=(k_0+\mu-iA_4,\vec{k})$. Then, the integration for four-momentum should be replaced as the following summation 
\begin{equation}
    \int\frac{d^4k}{(2\pi)^4}\rightarrow\int_T\frac{d^4k}{(2\pi)^4}\equiv iT\sum_n\int\frac{d^3\vec{k}}{(2\pi)^3}.
\end{equation}
One can see that the bubble diagram does not have a dependence on the external momentum $\textbf{q}$, and therefore the only nonzero contribution to the renormalization factor in Eq.~(\ref{1PI}) comes from the loop function $A(q^2)$.

By the following trick, we can further isolate the part of $A(q^2)$ that depends on the external momentum. To that end, let us rewrite this loop function as
\begin{equation}
\begin{split}
    A(q^2)=&\frac{1}{3}\Tr_c\int_T\frac{d^4k}{i(2\pi)^4}\frac{4[k^2+k\cdot q+M^2(\sigma)]}{[(k+q)^2-M^2(\sigma)][k^2-M^2(\sigma)]}
    \\
    =&\frac{1}{3}\Tr_c\int_T\frac{d^4k}{i(2\pi)^4}\frac{4[(k+q)^2-M^2(\sigma)]-2[q^2-4M^2(\sigma)]-2[q^2+2k\cdot q]}{[(k+q)^2-M^2(\sigma)][k^2-M^2(\sigma)]}.
\end{split}
\end{equation}
The last term in the numerator can be rewritten as $2[(k+q)^2-M^2(\sigma)]-2[k^2-M^2(\sigma)]$, and then if we change the internal variables to $d^4k\rightarrow d^4(k-q)$, the integration of these two terms will cancel. Then, the final result of the loop function is
\begin{equation}\label{loopA}
    A(q^2)=\frac{4}{3}\Tr_c\int_T\frac{d^4k}{i(2\pi)^4}\frac{1}{k^2-M^2(\sigma)}-2[q^2-4M^2(\sigma)]I(q,\sigma),
\end{equation}
where $I(q,\sigma)$ is the generating function of loop integral 
\begin{equation}
    I(q,\sigma)=\frac{1}{3}\Tr_c\int_T\frac{d^4k}{i(2\pi)^4}\frac{1}{[(k+q)^2-M^2(\sigma)][k^2-M^2(\sigma)^2]}.
\end{equation}
The first term in Eq.~(\ref{loopA}) is independent of external momentum and would not contribute to the renormalization factor. So, we only need to consider the second term, and then we can express the renormalization factor as 
\begin{equation}
    Z_\sigma^{-1}=18(1-\frac{G_D\sigma}{4G_S^2})^2[I(0,\sigma)+4M^2(\sigma)\frac{dI(0,\sigma)}{d\vec{q}^2}].
\end{equation}
In the following, we will deal with the integral of the finite-temperature loop function $I(q,\sigma)$.

\subsubsection{Calculation of $I(q,\sigma)$}

To incorporate the effect of temporal background $A_0=iA_4$, we define the generalized chemical potential as $\hat{\mu}=\mu+iA_4$ and recover $k_0=i(\omega_n-i\hat{\mu})$ with $\omega_n=(2n+1)\pi T$. Now, the integration becomes
\begin{equation}
    I(q,\sigma)=\frac{1}{3}\Tr_c\int\frac{d^3\vec{k}}{(2\pi)^3}\sum_{n=-\infty}^\infty\frac{1}{([(2n+1)\pi T-i\hat{\mu}-iq_0]^2+E^2_{k+q})([(2n+1)\pi T-i\hat{\mu}]^2+E^2_k)},
\end{equation}
where $E_k^2=\vec{k}^2+M^2(\sigma)$. Since the renormalization factor is computed when $q_0\rightarrow 0$, we can take this limit before implementing summation over $n$, giving the result
\begin{equation}
\begin{split}
    I(0,\vec{q},\sigma)=&\frac{T}{N}\Tr_c\int\frac{d^3\vec{k}}{(2\pi)^3}
    \\
    \times&\frac{E_k[{\rm Tanh}(\frac{\hat{\mu}-E_{k+q}}{T})-{\rm Tanh}(\frac{\hat{\mu}+E_{k+q}}{T})]-E_{k+q}[{\rm Tanh}(\frac{\hat{\mu}-E_{k}}{T})-{\rm Tanh}(\frac{\hat{\mu}+E_{k}}{T})]}{4TE_kE_{k+q}(E_{k+q}+E_k)(E_{k+q}-E_k)}.
\end{split}
\end{equation}
It is convenient to define the effective distribution functions~\cite{Kang:2022jbg,Kang:2024xqk} $F_+$ and $F_-$ as
\begin{equation}
    F_+(\omega)=\frac{2e^{-\frac{\omega-\hat{\mu}}{T}}}{1+e^{-\frac{\omega-\hat{\mu}}{T}}},\quad\quad\quad\quad F_-(\omega)=\frac{2e^{-\frac{\omega+\hat{\mu}}{T}}}{1+e^{-\frac{\omega+\hat{\mu}}{T}}}.
\end{equation}
Then the loop function can be expressed in terms of them
\begin{equation}
\begin{split}
    I(0,\vec{q},\sigma)=\frac{\Tr_c}{3}\int\frac{d^3\vec{k}}{(2\pi)^3}&\frac{1}{4E_kE_{k+q}}\Bigg[\frac{F_+(E_{k+q})-F_+(E_k)+F_-(E_{k+q})-F_-(E_k)}{E_{k+q}-E_k}
    \\
    &+\frac{2-F_+(E_{k+q})-F_-(E_{k+q})-F_+(E_k)-F_-(E_k)}{E_{k+q}+E_k}\Bigg],
\end{split}
\end{equation}
where the trace of the color subspace only acts on this effective distribution function.

In the $SU(3)$ gauge theory, the background gauge field, in the Polyakov gauge, is written as the diagonal phase matrix in Eq.~(\ref{BGF}). As a result, we can explicitly implement the trace over the effective distribution functions to get two functions of the PL, $f_{\pm}(\omega,l)$:
\begin{equation}
\begin{split}
   f_+(\omega,l)\equiv \frac{\Tr_c}{3}F_+(\omega)&=\frac{e^{-\Omega_+}}{3}\Tr_c\Bigg(\begin{matrix}
e^{i\theta_1}&0&0&\\
0&e^{i\theta_2}&0&\\
0&0&e^{i\theta_3}&
\end{matrix}\Bigg) \Bigg(\begin{matrix}
\frac{1}{1-e^{-\Omega_+}e^{i\theta_1}}&0&0&\\
0&\frac{1}{1-e^{-\Omega_+}e^{i\theta_2}}&0&\\
0&0&\frac{1}{1-e^{-\Omega_+}e^{i\theta_3}}&
\end{matrix}\Bigg)
\\
&=\frac{(l^*+2le^{-\Omega_+})e^{-\Omega_+}+e^{-3\Omega_+}}{1+3(l+l^*e^{-\Omega_+})e^{-\Omega_+}+e^{-3\Omega_+}},
\\
f_-(\omega,l)\equiv\frac{\Tr_c}{3}F_-(\omega)&=\frac{(l+2l^*e^{-\Omega_-})e^{-\Omega_-}+e^{-3\Omega_-}}{1+3(l^*+le^{-\Omega_-})e^{-\Omega_-}+e^{-3\Omega_-}},
\end{split}
\end{equation}
where $\Omega_\pm\equiv(\omega\pm\mu)/T$. In the last line, we have assumed that the background gauge field is a constant field and then substituted Eq.~(\ref{BGF}) into Eq.~(\ref{define:Pol}) to get this expression. Eventually, we can now express the loop function as a complicated function of $\sigma$, PL, and the chemical potential
\begin{equation}
\begin{split}
    I(0,\vec{q},\sigma)=\int\frac{d^3\vec{k}}{(2\pi)^3}&\frac{1}{4E_kE_{k+q}}\Bigg[\frac{f_+(E_{k+q},l)-f_+(E_k,l)+f_-(E_{k+q},l)-f_-(E_k,l)}{E_{k+q}-E_k}
    \\
    &+\frac{2-f_+(E_{k+q},l)-f_-(E_{k+q},l)-f_+(E_k,l)-f_-(E_k,l)}{E_{k+q}+E_k}\Bigg]
\end{split}
\end{equation}
From this expression, after a simplification, we can derive $I(0,\vec{q}=0,\sigma)$ and $dI(0,\vec{q},\sigma)/d\vec{q}^2|_{\vec{q}=0}$ as the following
\begin{equation}
\begin{split}
    I(0,\sigma)&=\int\frac{d\vec{k}^3}{(2\pi)^3}\frac{1}{4E_k^3}\bigg[1-f_+(\omega,l)-f_-(\omega,l)+E_k\bigg(\frac{\partial f_+(\omega,l)}{\partial\omega}+\frac{\partial f_-(\omega,l)}{\partial\omega}\bigg)\bigg]_{\omega=E_k}
    \\
    \frac{dI(0,\sigma)}{d\vec{q}^2}&=\int\frac{d\vec{k}^3}{(2\pi)^3}\frac{1}{16E_k^5}\bigg[3\bigg(f_+(\omega,l)+f_-(\omega,l)\bigg)-E_k\bigg(\frac{\partial f_+(\omega,l)}{\partial\omega}+\frac{\partial f_-(\omega,l)}{\partial\omega}\bigg)-1\bigg]_{\omega=E_k}.
\end{split}
\end{equation}
The resulting renormalization factor encounters $Z(\sigma)<0$ in some region of $\sigma$, indicating that the condensation is unstable and thus unphysic.

In computing $dI/d\vec{q^2}$, we have used the conclusion given by \cite{Reichert:2021cvs} that the Lorentz symmetry of zero-temperature theory required replacing $dI/d\vec{q^2}$ by $dI/d\vec{q^2}+\int\frac{d^3\vec{k}}{(2\pi)^3}\frac{1}{8E_k^5}$ to guarantee the positive definiteness of the renormalization factor. After (numerically) integrating the spatial momentum, we are left with a wave function factor $Z(\sigma,\mu,l;T)$. 
 
\subsection{Dark-QCD phase transition in the early universe}

The first thing to note is that, when we discuss the cosmic dark-QCD phase transition, it refers to the critical behavior in the dark-QCD plasma with a temperature $T$. It may or may not be in thermal equilibrium with the plasma of the standard model (SM), which has its own temperature $\mathcal {T}=\zeta T$ with $\zeta$ a free parameter that describes the ratio of two temperatures. In this paper, we only consider two limiting cases of interest, the dark-QCD dominant case $\zeta\sim 0$ and the thermal equilibrium case $\zeta=1$.

 \subsubsection{Routine parameters for the cosmic FOPT}

There are no available public codes to solve bounce equation Eq.~(\ref{bounce}) where a new term is present due to $Z_\sigma'\neq 0$, which is absent in the usual situations. In the mechanics analogy, this term results in an obstructing or accelerating proportional to the squared velocity, which further complicates the solution. We solve it using our own program based on the conventional shooting and overshooting method~\footnote{ The source code can be found on GitHub https://github.com/bhhua/VacuumTunneling. The solution and Euclidean action can also be found using the altered tunneling potential method, which can be found in the Appendix~\ref{TunnelingPM}. The Euclidean action $S_3/T$ obtained from the altered tunneling potential method is approximately 0.85 times of the exact numerical solution.}. With the solution, one can obtain the Euclidean action $S_E(T,\mu)$ and then calculate three routine phase transition parameters, $T_*$, $\wt\beta$ and $\alpha$.

The first parameter $T_*$ is the complete temperature of the phase transition, which is identified with the nucleation temperature $T_n$ or the percolation temperature $T_p$. They are always very close to the critical temperature and approximately equal to each other for the chiral phase transition. So, in this paper, we will only compute $T_n$, simply determined by the relation $S_E(T_n,\mu)/T_n\sim 140$. The second parameter $\wt\beta$ is the timescale of the phase transition, which can be computed by
\begin{equation}
    \wt\beta=-\frac{1}{H}\frac{d}{dt}\bigg(\frac{S_E}{T_n}\bigg)_{t=t_n}=T_n\frac{d}{dT}\bigg(\frac{S_E}{T_n}\bigg)_{T=T_n},
\end{equation}
where the second equality is because we only discuss the thermal equilibrium or dark-QCD dominated case thus $dT/dt=-TH$. The last parameter $\alpha$ is the strength of the phase transition, which can be defined by
\begin{equation}
    \alpha=\frac{\Delta \epsilon}{\rho_r(T_n)}
\end{equation}
where $\Delta\epsilon$ is the energy difference between two vacua  $\Delta\epsilon=\Delta V(T_n)-T_n\frac{d\Delta V(T_n)}{dT_n}$ with $\Delta V=V(\sigma_{f})-V(\sigma_{t})$ and $\rho_r$ is the radiation energy density of the universe
\begin{equation}
    \rho_r(T_n)=\rho^{DQCD}_r(T_n)+\rho^{SM}_r(T_n)=\frac{\pi^2}{30}(g^*_{DQC}+g^*_{SM}\zeta^4)T_n^4
\end{equation}
where $g_*^{QCD}=N^2-1+12N_f$ and $g_*^{SM}$ is the SM relativistic degree of freedom. Let us define the effective relativistic degree of freedom as $g_*=g^*_{DQC}+g^*_{SM}\zeta^4$ for convenience.

Now, let us treat the critical temperature $T_c$ and the chemical potential $\mu$ as free parameters to compute the above three phase transition parameters. In Table~\ref{table-2}, taking several groups of $\mu$ and $T_c$, we list the resulting phase transition parameters corresponding to the benchmark point B of the $SU(3)$ PNJL model given in Table~\ref{table-1}. In this table, we only collect the thermal equilibrium parameters. For the dark-QCD dominated case, the only difference is $g_*\rightarrow g_*^{DQCD}$ and $\alpha\rightarrow g_*\alpha/g_*^{DQCD}$. Here, for convenience in predicting GW, we use $T_c$ as the free parameter instead of $T_0$.
\begin{table} 
\begin{tabular}
{ |p{2.5cm}||p{1.8cm}|p{1.8cm}|p{1.8cm}|p{1.8cm}|p{1.8cm}|p{1.8cm}|p{1.8cm}|   } 
 \hline
 \multicolumn{8}{|c|}{Phase Transition Parameters of Benchmark B} \\
 \hline
Benchmark& $\mu/T_0$&$T_0$&$T_c$&$T_n$ &$g_*$& $\alpha$&$\wt\beta$\\
 \hline
B& 1.421&1.4695$T_c$&10GeV&0.9992$T_c$&138&0.0234&791545\\
B& 1.65&1.9309$T_c$&10MeV&0.99348$T_c$&53&0.5628&51083\\
B& 1.65&1.9309$T_c$&100MeV&0.9948$T_c$&101&0.2953&51083.0\\
B& 1.65&1.9309$T_c$&1GeV&0.9948$T_c$&129&0.2312&51083.0\\
B& 1.65&1.9309$T_c$&10GeV&0.9948$T_c$&138&0.2167&51083.0\\
B& 1.65&1.9309$T_c$&100GeV&0.9948$T_c$&151&0.1976&51083.0\\
B& 1.8&3.6122$T_c$&10MeV&0.9261$T_c$&53&2.2794&2901.27\\
B& 1.8&3.6122$T_c$&100MeV&0.9261$T_c$&101&1.1961&2901.27\\
B& 1.8&3.6122$T_c$&1GeV&0.9261$T_c$&129&0.9365&2901.27\\
B& 1.8&3.6122$T_c$&10GeV&0.9261$T_c$&138&0.8754&2901.27\\
B& 1.8&3.6122$T_c$&100GeV&0.9261$T_c$&151&0.8000&2901.27\\
\hline
\end{tabular}
\caption{The phase transition parameter for the Benchmark points B at different chemical potential and the critical temperature.}  
\label{table-2}
\end{table}

\subsubsection{Dynamic roles of chemical potential in the cosmic chiral phase transition}

From Table~\ref{table-2}, we find that the nucleation temperature $T_n$ for the chiral phase transition is very close to the critical temperature $T_c$. This means that the $\alpha$ parameter is mainly contributed by latent heat $Td\Delta V/dT$ at $T_n$. It is observed that increasing the chemical potential leads to the following consequences of phase transition parameters:
\begin{itemize}
    \item $T_n$ decreases. 
    \item $\wt\beta$ also decreases. 
      \item $\alpha$ increases for fixed $T_c$. 
\end{itemize}
Such features are conducive to the generation of observable GWs.

\begin{figure}[htbp]
\centering 
\includegraphics[width=0.45\textwidth]{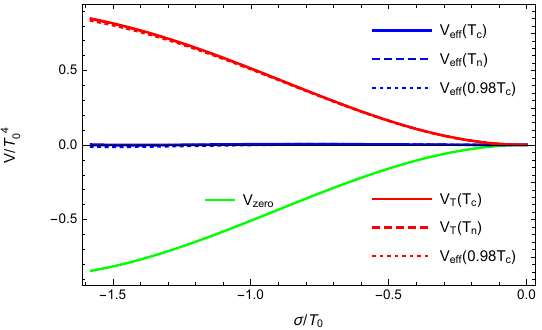}
\includegraphics[width=0.45\textwidth]{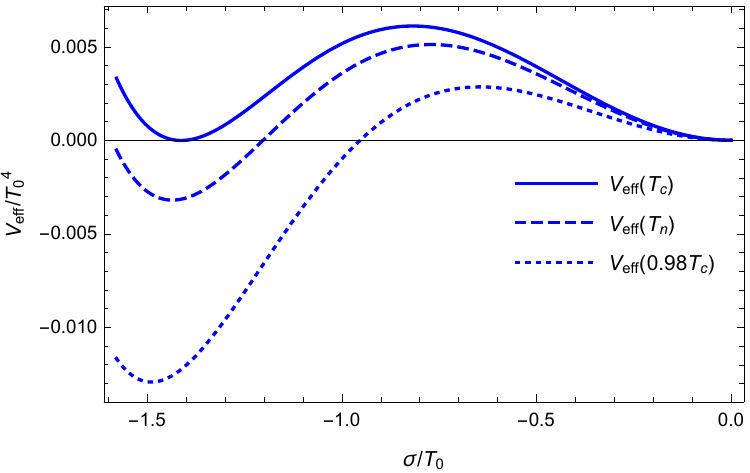}
\caption{The left panel shows the fine balance structure of the zero and finite effective potential for $G_s=2,2/\Lambda^2$ and $\mu=1.65 T_0$. The right panel enlarged the total effective potential, which is shown as the blue line in the left panel. The solid line shows the result from $T=T_c$, the dashed line shows the result from $T=T_n\approx 0.995 T_c$ and the dotted line shows the result from $T=0.98T_c$.}\label{VEFFDEOMO}
\end{figure}

Let us explain the possible reasons for the above changes. It is more appropriate to start by understanding why the $\wt\beta$ parameter is quite large in the QCD-like phase transition , e.g., in the confinement phase transition  in pure-Yang Mills theory~\cite{Kang:2022jbg,Huang:2020crf,Halverson:2020xpg}. 
We speculate that this is attributed to the dramatically varying potential as $T$ close to $T_c$. The reason for this behavior is that the zero temperature effective potential $V_{zero}$ and the finite-temperature effective potential $V_T$ demonstrate a fine balance at $T=T_c$. More precisely, both $V_{zero}$ and $V_T$ are quite large, but their contributions to the total effective potential $V_{eff}$ are opposite; around $T_c$, those two large parts cancel each other out and just leave a small degenerate vacuum structure. Consequently, $T_c$ is dynamically tuned, and as $T$ drops further a little, the fine balance behavior will be heavily broken and $\Delta V(T)=V(\sigma_t,T)-V(\sigma_f,T)$ will change dramatically with temperature. The above features are manifest in Fig.~\ref{VEFFDEOMO}, where we take three very close ($\sim {\cal O}(1\%)$) temperatures around $T_n$ , $T=T_c$, $T_n$ and $0.98 T_c$: for three cases, the curves of $V_T$ almost overlap (left panel), but $V_{eff}(\sigma)$ drops very fast around the chiral symmetry breaking vacuum, as seen in the right panel. The resulting $\tilde{\beta}$ is large, which may be understood in the thin wall limit, where $S_3(T)\propto 1/\Delta V(T)$~\cite{Coleman:1977py} and therefore $\wt \beta=T\frac{d}{dT}\left(\frac{S_3}{T}\right)\sim -\frac{1}{\Delta V(T)^2}\frac{d\Delta V(T)}{dT}$, which is enhanced by $-d\Delta V(T)/dT\gg 1$.



However, in the presence of a large chemical potential $\mu$ within the effective potential, the dependence on $T$ is weakened by this large energy scale. Then, to cause the same variation of the effective potential, $T_c-T$ should become larger to counter the effect of $\mu$, which thus decreases $T_n$ and $\wt\beta$ simultaneously. Given that we took a fixed $T_0$, this would lead to fewer latent heat releases, as discussed in Section~\ref{PDiagram}. However, when demonstrating the impacts of increasing $\mu$ on the phase transition  parameters in Table~\ref{table-2}, we take $T_c$ (the critical temperature of dark-QCD) rather than $T_0$ (the intrinsic scale of the PYM) as fixed. Then, to maintain $T_c$ with increased $\mu$, $T_0$ should increase at the same time, indicating a PYM with a higher energy scale, thus giving a larger $\alpha$ as expected.


To end with this section, let us mention that the dynamical tuning around $T_c$ thus a large $\wt\beta$ may also occur in electroweak-like phase transitions. We analyze the strange FOPT considered in the toy model given in Subsection~\ref{EW:ex}, and find that this is also caused by the dynamical adjustment between the effective potentials of zero and the finite temperature. 





\section{Enhanced GW Signal from large chemical potential}

\subsection{The GW sources and spectra}

It is well known that, during cosmic FOPT, there are three types of sources for GWs, i.e. bubble collision~\cite{Kosowsky:1992vn,Jinno:2016vai,Jinno:2017ixd}, sound wave~\cite{Hindmarsh:2013xza,Hindmarsh:2015qta}, and magnetohydrodynamic (MHD) turbulence~\cite{Kahniashvili:2009mf,RoperPol:2019wvy}. The latent heat released during the FOPT is distributed in these sources, but the fractions of each source are not yet precisely determined, and one can only estimate them by the mean field-hydrodynamical system. Usually, the fraction of latent heat transferred to the bubble collision is believed to be negligible $\kappa_{col}\ll 1$, provided that $\alpha$ does not become extremely large~\cite{Ellis:2019oqb}.

The GW sources are dominated by two bulk motions of the plasma. One bulk motion is the sound wave propagating in the plasma after the percolation, and the latent heat that goes into it is estimated to be~\cite{Espinosa:2010hh}
\begin{equation} 
\begin{split}
    &\kappa_{sw}\approx \alpha(0.73+0.083\sqrt{\alpha}+\alpha)^{-1}\quad\quad\quad\quad\quad\quad v_w\sim 1,
    \\
    &\kappa_{sw}\approx v_w^{\frac{6}{5}}6.9\alpha(1.36-0.037\sqrt{\alpha}+\alpha)^{-1}\quad\quad\quad v_w\leq 0.1,
\end{split}
\end{equation}
where $v_w$ is the bubble wall velocity, and we will comment it specifically later. The peak frequency of the GW at $T_n$ is $f_{sw,*}=2(8\pi)^{1/3}/[\sqrt{3}(v_w-c_s)R_*]$ with $c_s=1/\sqrt{3}$ the speed of sound in the plasma and $R_*$ the average bubble separation at the collision. This separation is related to the typical time scale of phase transition  via $R_*=(8\pi)^{1/3}v_w/\beta_n$, where $\beta_n$ is the duration of phase transition($\beta$) at nucleation temperature, given the exponential approximation of $\Gamma(T)$ around $T_n$. The observed GW spectra today peaks at $f_{sw}=f_{sw,*}a_0/a(T_n)$, as a result of the redshift from the GW production time $t_n$ to today. 
Then the  peak frequency is parameterized as \cite{Ellis:2019oqb}
\begin{equation} \label{SW} 
\begin{split}
f_{sw}&=1.65\times 10^{-5}\L\frac{T_n}{100 \rm GeV}\R\L\frac{g_*}{100}\R^{\frac{1}{6}} \frac{3.4}{(v_w-c_s)H_* R_*} \rm Hz
\\
&=2.75\times10^{-5}\frac{\wt\beta}{v_{w}}\L\frac{T_n}{100 \rm GeV}\R\L\frac{g_*}{100}\R^{\frac{1}{6}} \rm Hz.
\end{split}
\end{equation}
The amplitude of the GW spectrum for the sound wave is then given by~\cite{Guo:2020grp}
\begin{align} 
h^2\Omega_{sw}(f)&=6.3\times 10^{-5}\f{1}{\wt\beta^2}\L\frac{ \kappa_{sw}\alpha}{1+\alpha}\R^{2}\L\frac{100}{g^*}\R^{\frac{1}{3}}v_{w}^2\frac{f^3}{f_{sw}^3}\left[\frac{7}{4+3(f/f_{sw})^2}\right]^{\frac{7}{2}}.
\end{align}
When the sound wave period ends and the fluid flow becomes non-linear, the MHD turbulence is generated. For $\tau_{sw}$ longer than one Hubble time scale, this bulk motion is suppressed, having an efficiency factor $\kappa_{turb}\sim0.05\kappa_{sw}$~\cite{Hindmarsh:2015qta}. The GW spectrum from MHD turbulence is given by~\cite{Caprini:2009yp},
\begin{equation}\label{TURB}
h^2\Omega_{turb}(f)=3.35\times10^{-4}\f{1}{\wt\beta}\L\frac{ \kappa_{sw}\alpha}{1+\alpha}\R^\frac{3}{2}\L\frac{100}{g}\R^{\frac{1}{3}}v_{w} \frac{(f/f_{turb})^3}{[1+(f/f_{trub})]^{\frac{11}{3}}(1+8\pi f/h)},
\end{equation} 
which, compared to $\Omega_{sw}(f)$, shows a moderately large suppression $\sim{\cal O}{(10)}$ in the high-frequency region. The peak frequency is similar to that of the sound wave source~\cite{Ellis:2019oqb}, 
\begin{equation} \label{}
\begin{split}
f_{turb}&=1.65\times 10^{-5}\L\frac{T_n}{100 \rm GeV}\R\L\frac{g_*}{100}\R^{\frac{1}{6}} \frac{3.9}{(v_w-c_s)H_* R_*} \rm Hz
\\
&=3.15\times10^{-5}\frac{\wt\beta}{v_{w}}\frac{T}{100 \rm GeV}\L\frac{g}{100}\R^{\frac{1}{6}}\rm Hz.
\end{split}
\end{equation}

Although the above formulas are empirical formulas given by simulation, some recent studies have found that the GWs given by them are not accurate~\cite{Guo:2023gwv,Guo:2024gmu}. However, those new results come from the ideal fluid systems, and due to the complexity of the numerical simulation of the QCD interacting fluid system, we will still use those formulas to compute the gravitational wave signal from the chiral phase transitions. 

\subsection{A comment on the bubble wall velocity of dark-QCD phase transition} 

The velocity of the bubble wall is an important parameter in the prediction of GW~\cite{Ai:2021kak,Ai:2023see}. In the electroweak-like phase transition, the $v_w$-dependent plasma friction acting on the wall is clearly attributed to mass change of the plasma particles after passing through the wall. So, we can calculate $v_w$ in a relatively confirmative way. However, for the confinement phase transition, we do not have such a picture (essentially, this is due to ignorance of the dynamic role of the wall in the gluon-to-glueball process), and hence there is no well-accepted approach to determine the bubble velocity. It is estimated to be around ${\cal O}(0.1)$ via different approaches~\cite{Kajantie:1986hq,Ignatius:1993qn,Bea:2021zsu,Bigazzi:2021ucw,Kang:2024xqk}, for example, treating the QGP as a pool of quasiparticles moving in the background $A_4$~\cite{Kang:2024xqk} or using the large jump in degrees of freedom of the fluid system, which may be suitable for studying the confinement phase transition~\cite{Sanchez-Garitaonandia:2023zqz}, even using the holographic method~\cite{Bea:2021zsu,Bigazzi:2021ucw}.

For the chiral phase transition, the quarks gain a constituent mass from the $\sigma$ condensate, and one may speculate that in the NJL model the wall velocity can be determined as the electroweak-like phase transition. However, the PL value also changes along the bubble wall, indicating that some partons confine in the course of wall passing, which may induce a friction force. This again makes the prediction of $v_w$ difficult, and it depends on how to incorporate this confinement effect. It is beyond the scope of this paper and we will take $v_w\sim 1$ to estimate the largest GWs and only comment on the $v_w\sim 0.1$ case.

\subsection{Prospect of GW from chemical potential driven dark-QCD phase transition}

This subsection is devoted to demonstrating the prospect of a stochastic GW background generated by the cosmic dark-QCD phase transition driven by a chemical potential. For a given bubble wall velocity, GW is mainly shaped by $G_S$ and $\mu$ and the dynamic intrinsic scale $T_0$ (or $T_c$), and we will analyze their impacts separately. 


\begin{figure}[htbp]
\centering 
\includegraphics[width=0.45\textwidth]{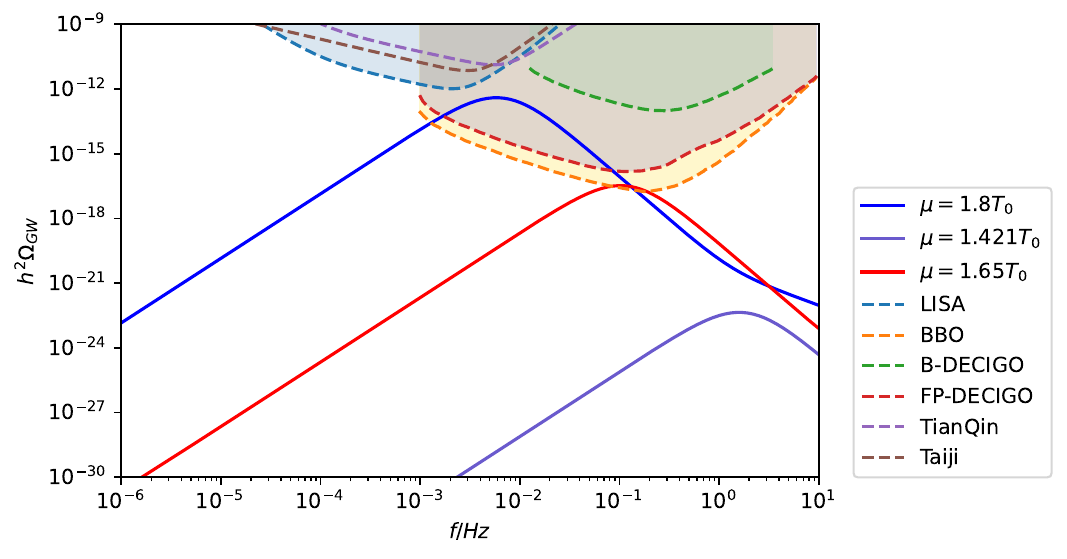} 
\includegraphics[width=0.45\textwidth]{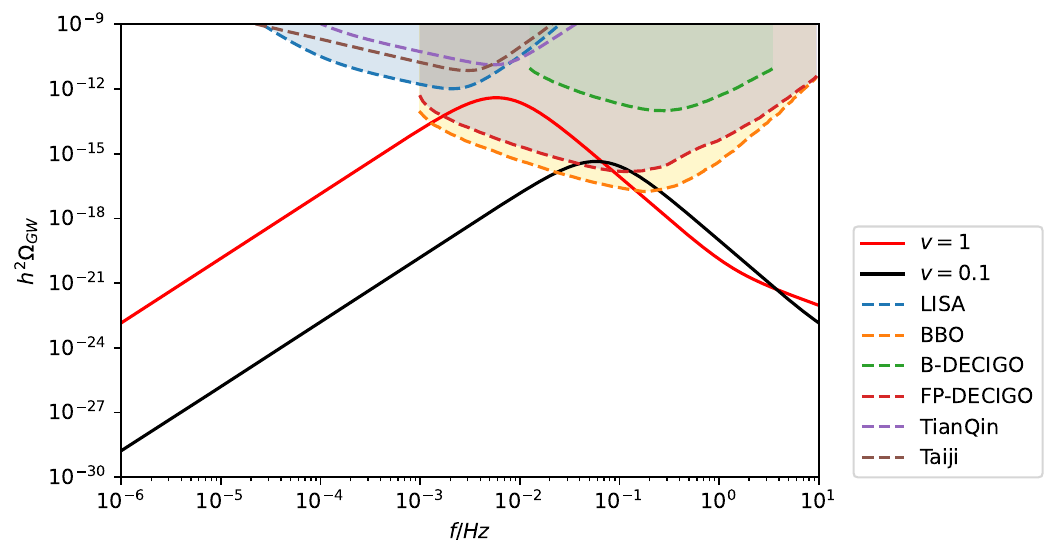} 
\caption{$G_S=2.2/\Lambda^2$, first is chemical potential from $\mu=1.8T_0$, $1.65T_0$ and $1.421T_0$, second is velocity from $v=1$ to $0.1$ at $T_c=10{\rm GeV}$ with $\mu=1.8T_0$.} \label{GW1}
\end{figure}
\begin{figure}[htbp]
\centering 
\includegraphics[width=0.45\textwidth]{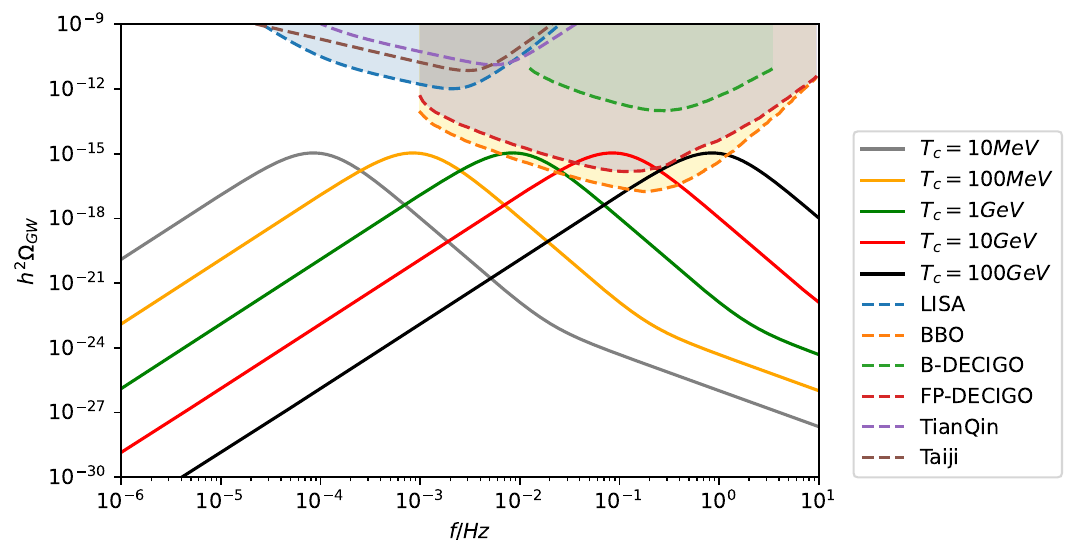} 
\includegraphics[width=0.45\textwidth]{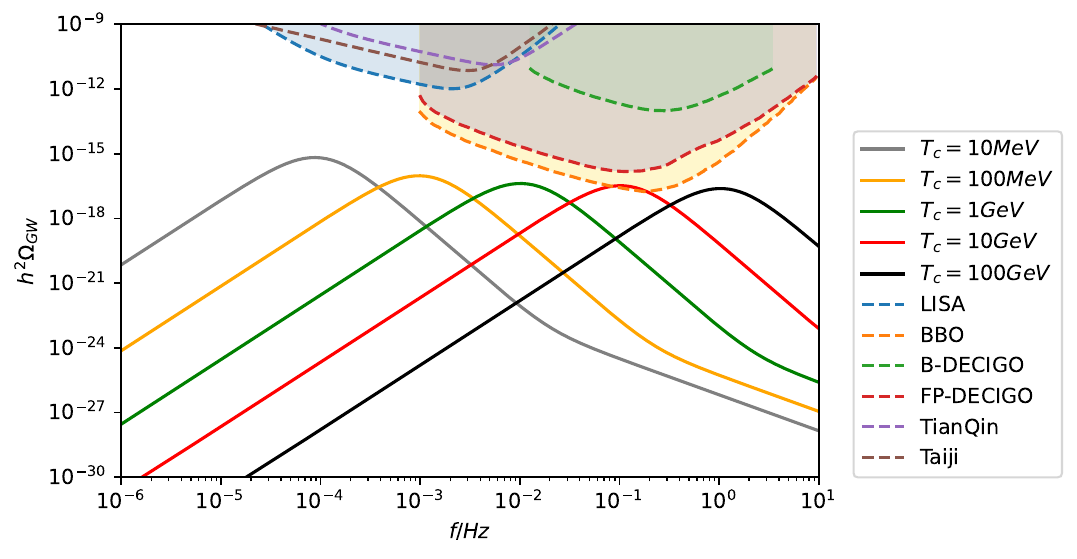} 
\includegraphics[width=0.45\textwidth]{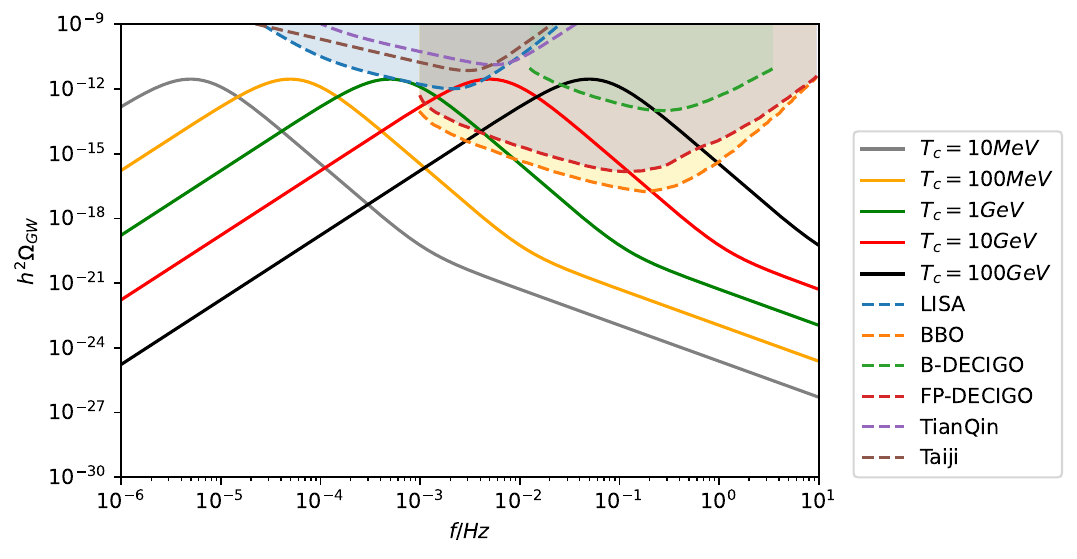} 
\includegraphics[width=0.45\textwidth]{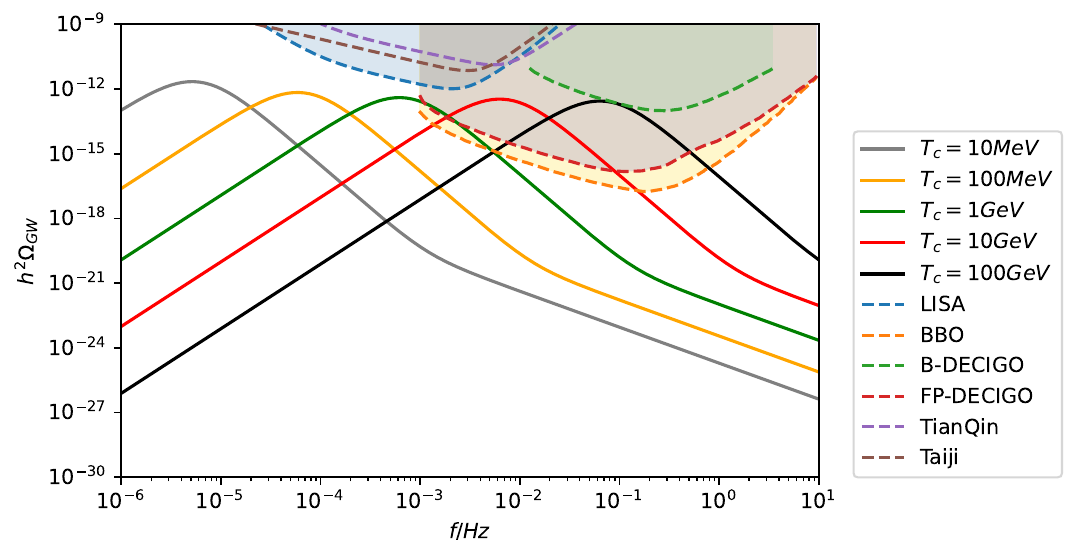}
\caption{Top: GWs for $(G_S=2.2/\Lambda^2,\mu=1.65T_0)$ with several critical temperatures; bottom:  the same for $(G_S=2.2/\Lambda^2,\mu=1.8T_0)$. 
The left panels are for the dark-QCD-dominated case with $\zeta\to 0$ while the right panels zhelare for $\zeta\to 1$. } \label{GW2} 
\end{figure}

\begin{itemize}
    \item First, let us calculate the GWs with different chemical potentials, taking the phase transition parameters from Table~\ref{table-2} for $T_c=10~ \rm GeV$ and $G_S=2.2/\Lambda^2$. We show the GW signals in the left panel of Fig.~\ref{GW1} for three values $\mu/T_0=1.421$ (lower slate blue line), $\mu/T_0=1.65$ (middle pink line), and $\mu/T_0=1.8$ (upper blue line).  Notice that $\mu/T_0=1.421$ is very close to the critical point in the phase diagram in Fig.~\ref{phsdiag} while $\mu/T_0=1.8$ can be even larger. It is clearly seen that the GW signal is highly sensitive to $\mu$. For the $\mu$ increase for just 10$\%$, the peak frequency is red-shifted considerably, and the amplitude even increases by many orders of magnitude.
    \item In the previous plots, we have taken $v_w\sim 1$ to estimate the strongest GWs. Considering the uncertainty of the velocity of the bubble wall, we should estimate the prospect of GW for a relatively smaller $v_w$. In the right panel of Fig.~\ref{GW1}, we plot the GWs for $\mu/T_0=1.8$ with two typical bubble velocity, $v_w\sim 1\ \&\ 0.1$. One can see that, although the GW amplitude is substantially for the smaller $v_w$, the resulting peak frequency of the GW may still lie within the sensitivity curve of the Big Bang Observer (BBO) detector~\cite{Crowder:2005nr,Corbin:2005ny,Harry:2006fi}. So, a part of the parameter space can overcome moderately small velocity suppression to generate testable GWs. 
    \item 
    The critical temperature $T_c$ is directly related to the location of the peak frequency of the GWs, and the plots for different $T_c$ in Fig.~\ref{GW2} manifest this. The almost simple shift of the GW spectra with an increase in $T_c$ is well expected from the expression \eqref{SW} that shows $f_{sw}\propto T_c$, with other phase transition parameters $\widetilde \beta$ and $\alpha$ not sensitive to $T_c$ (only $\alpha$ receives a small decrease due to the increase in plasma degrees of freedom; see Table.~\ref{table-2}). 
\end{itemize}
To gain a more comprehensive expression on the GWs of the $\mu$-driven dark-QCD phase transition, in Fig.~\ref{GW2} we show two chemical potentials $\mu/T_0$ (1.65 for the top panels and 1.8 for the bottom panels) for two scenarios of dark-QCD temperature over SM temperature ($\zeta\to 0$ for left panels and $\zeta\to 1$ for right panels, with the former getting a slightly better chance.); again we take $v_w=1$. In general, the predicted signals that may be hunted fall in the intermediate frequency band of $10^{-3}-1$ Hz, which can be best covered by BBO but only marginally by other space detectors such as Lisa~\cite{Caprini:2015zlo,LISA:2017pwj,Caprini:2019egz}, Tianqin~\cite{TianQin:2015yph,TianQin:2020hid} and  Taiji~\cite{Ruan:2018tsw,Ruan:2020smc}, which are designed with the best sensitivity in the low frequency band. In turn, only dark-QCD with $100 {\rm GeV}\gtrsim T_c\gtrsim 1$ GeV has a promising chance of being detected, given a sufficiently higher $\mu$.   

\begin{figure}[htbp]
\centering 
\includegraphics[width=0.45\textwidth]{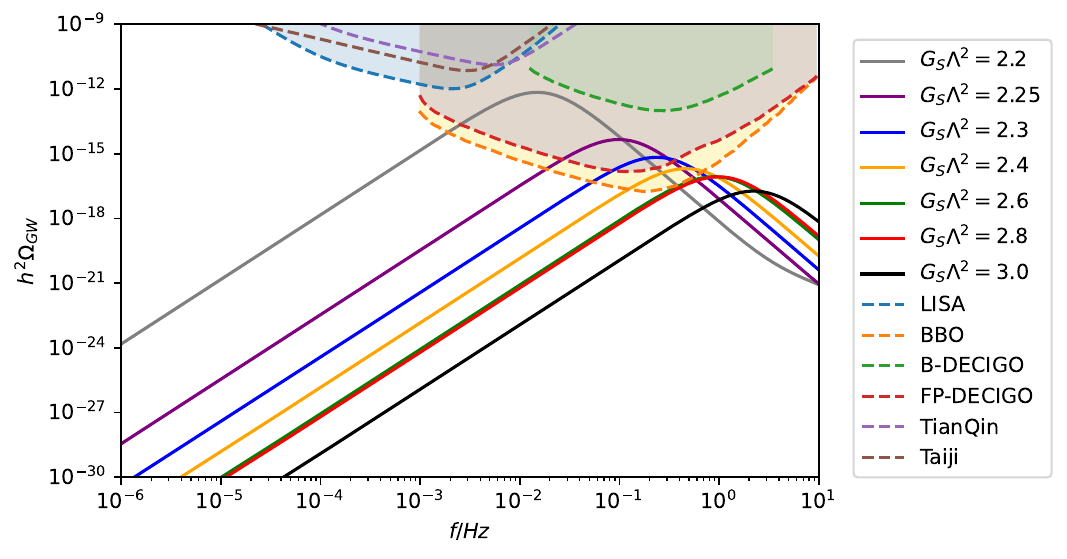} 
\includegraphics[width=0.45\textwidth]{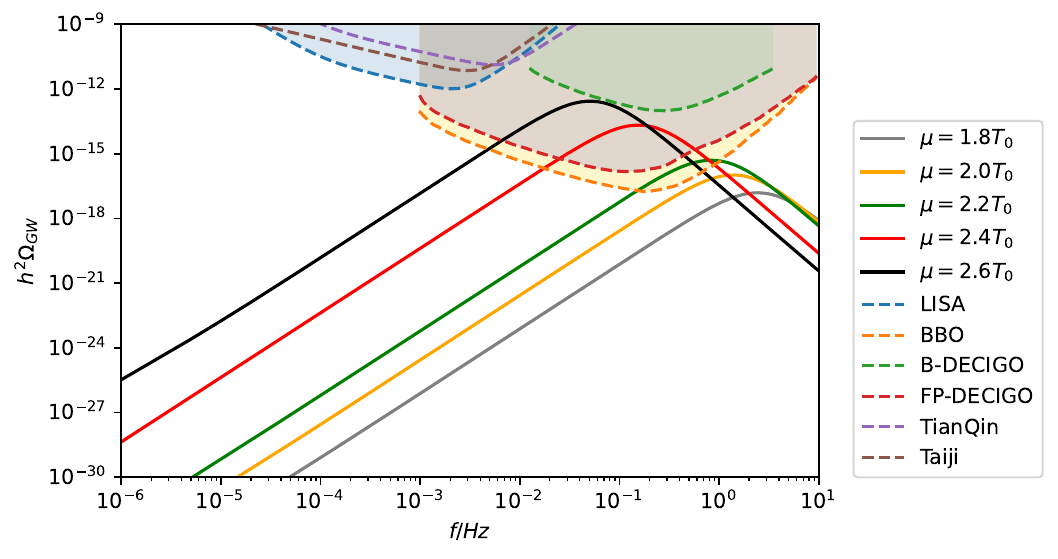} 
\caption{ Left: GWs from different $G_S$ with a fixed $\mu=1.8T_0$; right: GWs from the $G_S\Lambda^2=3$ with varying chemical potential. In both panels we take $T_0=100~ \rm GeV$ and $\zeta=1$. } \label{GW3} 
\end{figure}

Until now, we have not discussed the effect of $G_S$, which is the crucial parameter of chiral symmetry breaking in the NJL model. We show the GWs for different $G_S$ in the left panel of Fig.~\ref{GW3}, taking the same $T_0$ and chemical potential, to find that as $G_S$ increases, the resulting GWs shift to the higher frequency region, with amplitude obviously suppressed. This behavior can be simply understood as follows. A larger $G_S$ leads to a higher chiral symmetry breaking scale $\langle \sigma\rangle$ at zero temperature, so $V_{zero}$ has a deeper minimum. Following the previous argument of dynamical tuning, a larger $V_T(T=T_c)$ is necessary to cancel $V_{zero}(\langle \sigma\rangle)$, which then means a higher critical temperature $T_c$, which explains the movement of the peak frequency.   Recalling that $\mu$ is able to provide a large scale in $V_T$ to relax the dynamical tuning and produce a smaller $\beta$, then a larger $G_S$ would weaken the role of $\mu$ which is not increased simultaneously, giving rise to a much larger $\beta$.

A larger $G_S$ would accommodate a larger $\mu$ (see Fig.~\ref{phsdiag}), and thus we also increase $\mu$ correspondingly, beyond $1.8T_0$, to see the change of the resulting GWs; a sample for $G_S\Lambda^2=3$ with several increasing $\mu$ is given in the right panel of Fig.~\ref{GW3}. As $\mu$ increases to the almost maximal value in the phase diagram, $\mu=2.6T_0$, the resulting GWs become roughly similar to the one in the $(G_S\Lambda^2\sim 2.2,\mu=1.8T_0)$ case. In other words, although we do not explore the full parameter space over the $(G_S,\mu)$ plane, for a given $T_0$, by fixing $G_S\Lambda^2$ and roaming in the $\mu$ parameter space, we can also roughly understand the distribution of the GWs spectrum in the entire theory.

\section{conclusion and discussion}

It is known that within QCD, a large chemical potential may drive the chiral symmetry phase transition to be a first-order one. However, it does not work effectively to generate a promising GW signature in the early universe, due to the cosmological bound. In addition, how the chemical potential changes and affects the chiral phase transition, in particular the resulting dynamics of this phase transition, is not very clear. Therefore, it is of importance to make a profound study in the context of dark-QCD which is free of constraints and thus may provide a detectable GW source. A large chemical potential of the order of temperature may be generated via the AD mechanism. 

In this paper, we adopt the conventional PNJL model in the presence of a chemical potential to study the chiral phase transition of the three-flavor dark-QCD in the chiral limit, moreover with a vanishing $G_D$. We plot its phase diagram and find that the chiral phase transition is indeed first-order given a sufficiently large chemical potential. We try to uncover the reasons for this phase transition, as well as its features. It is found that the barrier is generated by the dynamical fine balance between the zero- and finite-temperature effective potential, which is similar to the case in the deconfinement-confinement phase transition in the hot pure Yang-Mills system. Hence, the time duration of the phase transition is fairly short, that is, giving rise to a large $\beta/H$, which is one of the reasons why the GW amplitude is suppressed in the QCD-like phase transition. However, we have shown that the relatively larger chemical potential, introducing a new large scale in the effective potential, helps to substantially decrease $\beta/H$. Moreover, increasing the chemical potential significantly enhances the discovery prospect of the GWs from the dark-QCD phase transition. As a consequence, for the transition that occurs at the critical temperature above 1 GeV and below 100 GeV, the produced GWs may reach the sensitivity of the future BBO detector. This is one of the main results of this paper.

The cosmic chiral phase transition raises a technique issue. We have to consider the wave function renormalization for the order parameter field $Z(\sigma)$, and then the conventional way to solve the bounce equation should be modified. We develop two numerical approaches to deal with this issue, to find good consistence. And a specific program will appear elsewhere. 



In this work, we only study the dark chiral phase in dark-QCD with color number $N=3$, and it is interesting to discuss the general case with $N>3$. As we discussed in Section~\ref{PTQCD}, the Polykov loop $L$ is a key parameter for the QCD phase transition. As a $N\times N$ matrix, it contains  multiple components and thus we need to construct some linear independent fields to describe the eigenvalue of $L$. For $N=3$, we only need two linear independent fields, which can be chosen as the trace of $L$ and its complex conjugate, $l$ and $l^*$. However, for $N=4$, we need to add a new field ${\rm tr} L^2$. This complication increases with $N$ and seems very difficult to handle for large $N$. In pure Yang-Mills theory, the assumption of equal eigenvalue of $L$ helps simplify the problem~\cite{Kang:2022jbg,Dumitru:2012fw}. However, adding dynamic quarks with chemical potential violates this nice property even for $N=3$. Therefore, it is nontrivial to study the cases with $N>3$. However, we still expect stronger GWs to be produced, as the latent heat release increases with $N$. In addition, exploring the effect of chemical potential in the electroweak-like phase transition is of interest but relatively few, and we notice that a work focusing on this has already appeared~\cite{Qin:2024idc}. We may leave these topics for future works.


\noindent {\bf{Acknowledgements}}

This work is supported in part by the  National Key Research and Development Program of China Grant No.2020YFC2201504 and National Science Foundation of China (11775086).

\appendix
\section{Effective potentials for the NJL model from path integral}\label{pathintegral}

The NJL model merely contains fermonic degrees of freedom, and they are interacting via the nonrenormalizable four-fermion terms which may arise after integrating out the gluons,
\begin{equation}\label{NJL}
\begin{split}
    \mathcal{L}_{NJL}=\bar{q}i\slashed D q+G_S\left[(\bar{q} q)^2+(\bar{q}i\gamma^5\tau_i q)^2\right].
\end{split}
\end{equation}
We consider two-flavor quarks with the covariant derivative $D_\mu=\partial_\mu-i A_\mu $ with $A_\mu$ the background gauge field $A_\mu=A_0\delta_{\mu 0}=iA_4\delta_{\mu 0}$, which is relevant around the critical temperature. To describe chiral symmetry spontaneously breaking and introduce mesons, one can turn to the bosonization approach, which introduces the auxiliary fields $\sigma$ and $\vec{\pi}$ with the following Gaussian integration
\begin{equation}
    Const=\int\mathcal{D}[\sigma]\mathcal{D}[\pi_i]\exp\left(i\int d^4x\frac{-1}{4G}\Bigg[(\sigma+2G\bar{q}q)^2+(\pi_i+2Gi\bar{q}\gamma_5\tau_iq)^2\Bigg]
    \right).
\end{equation}
It eliminates the two non-renormalized terms, and then the generating functional becomes 
\begin{equation}
    Z=\int\mathcal{D}[\sigma]\mathcal{D}[\pi_i]\mathcal{D}[q]\mathcal{D}[\bar{q}]\exp\left[i\int d^4x\mathcal{L}_{eff}(\sigma,\pi_i,q,\bar{q},A_\mu)\right],
\end{equation}
where the effective action is given by
\begin{equation}\label{NJL:B}
    \mathcal{L}_{eff}(\sigma,\pi_i,q,\bar{q},A_\mu)=\bar{q}[i\gamma^\mu D_\mu-\sigma-i\gamma_5\tau_i \pi_i]q-\frac{1}{4G}[\sigma^2+\pi_i^2].
\end{equation}

Now, the quark fields can be completely integrated out via Gaussian integration, leaving the bosonization of NJL to describe the collective fields
\begin{equation}
\begin{split}
 Z   &=\int\mathcal{D}[\sigma]\mathcal{D}[\pi_i]\exp\left(-\frac{i}{4G}\int d^4x[\sigma^2+\pi_i^2]+N_f{\rm tr}\log\Bigg[i\gamma^\mu D_\mu-\sigma-i\gamma_5\vec{\tau}\cdot\vec{\pi}\Bigg]\right),
\end{split}
\end{equation}
where the non-local functional determinate interaction incorporate the dynamics of $\sigma$ and $\pi$. For instance, calculating the wave-function renormalization factor and extracting the effective potential for the background of $\sigma\to \sigma_0$,
\begin{equation}
    V(\sigma)=\frac{\ln Z}{V}=\frac{\sigma_0^2}{4G}+N_f{\rm tr_c}\int\frac{d^4k}{(2\pi)^4}\log\Bigg(\det[\gamma^\mu (k_\mu-iA_\mu)-\sigma_0]\Bigg).
\end{equation}
For our purposes, the second term is calculated at finite temperature, on a $A_4$ background. Following the procedure~\cite{Quiros:1999jp}, it can be written as the following integral
\begin{equation}
\begin{split}
    V_{NJL}(\sigma,T)=\frac{\sigma_0^2}{4G}+2TN_f\sum_n{\rm tr_c}\int\frac{d^3\vec{k}}{(2\pi)^3}\log\Bigg[ (\omega_n-A_4)^2-|\vec{k}|^2-\sigma_0^2\Bigg],
\end{split}
\end{equation}
with $\omega_n=i(2n+1)\pi T$ for the canonical ensemble.

If the quark and anti-quarks have a large asymmetry, we need to introduce the chemical potential, and then the ensemble switches from canonical to grand canonical: $H\to H-\mu N$ with $N$ the corresponding number operator, or the conserved charge density. For fermions, $N=\bar q \gamma^0 q$, which is independent of momentum, this amounts to modifying the generating functional by adding a term $\mu\bar q \gamma^0 q$ to the Lagrangian~\cite{Kapusta:2006pm},
\begin{equation}
\begin{split}
    \mathcal{L}_{\mu-NJL}=\bar{q}i[\slashed \partial+  \gamma^0(  iA_4-\mu)]q+G_S\left[(\bar{q} q)^2+(\bar{q}i\gamma^5\tau_i q)^2\right].
\end{split}
\end{equation}
That is, the temporal background $iA_4$ plays the same role as the real chemical potential, which explains why $A_4$ is dubbed the imaginary chemical potential. As a consequence, one can simply get $\ln Z$ by replacing $\mu\to \mu-iA_4$ in the one given in~\cite{Kapusta:2006pm} (an alternative derivation is given in \cite{Kang:2022jbg}), to give 
\begin{equation}
\begin{split}
    V_{NJL}(\sigma,T)&=\frac{\sigma_0^2}{4G}-2N_cN_f\int\frac{d^3\vec{k}}{(2\pi)^3}E_k
    \\
    &+2TN_f{\rm tr_c}\int\frac{d^3\vec{k}}{(2\pi)^3}\Bigg(\log\Bigg[ 1+e^{-\beta(E_p-\mu+iA_4)}\Bigg]+\log\Bigg[ 1+e^{-\beta(E_p+\mu-iA_4)}\Bigg]\Bigg)
    \\
    &=\frac{\sigma_0^2}{4G}-2N_cN_f\int\frac{d^3\vec{k}}{(2\pi)^3}E_k
    \\
    &+2TN_f{\rm tr_c}\int\frac{d^3\vec{k}}{(2\pi)^3}\Bigg(\log\Bigg[ 1+\hat{L}e^{-\beta (E_p-\mu)}\Bigg]+\log\Bigg[ 1+\hat{L}^\dagger e^{-\beta (E_p+\mu)}\Bigg]\Bigg)
\end{split}
\end{equation}
 where the first term is the tree-level potential \eqref{treep}, the second term is the vacuum energy, which is proportional to the cut-off scale \eqref{vacengy}, and the third term is just the normal thermal function of the chiral and Polyakov background fields \eqref{NJL:T}. The above derivation can be generated for the case with more flavors. For example, $N_f=3$ can be found in~\cite{Kerbikov:1995pa}.

\section{The meson spectrum in the dark-QCD in the chiral limit}\label{spectrum}

In this appendix we calculate the  mass sepctrum and decay constants for the mesons in the chiral limit. The meson mass can be obtained by matching the scattering amplitude for some light-quark $i\mathcal{M}^{qq}_{\rm NJL}$ calculated in the original NJL model \eqref{NJL} to $i\mathcal{M}^{qq}$, obtained in the model \eqref{NJL:B} which replaces the four fermion terms with the mesons. They are respectively proportional to $\frac{N(p^2)}{D(p^2)}$ and $\frac{1}{p^2-m^2}$, with $m$ the meson mass to be determined. Then, $D(p^2)\propto p^2-m^2$ as long as $N(p^2)$ is not divergent. Therefore, the meson masses are identified with the pole of $D(p^2)$. For $a=\sigma,\pi$, their masses are solutions of the following equations~\cite{Kohyama:2016fif}
\begin{equation}
\begin{split}
    &1-2K_a\Pi_a(p^2)|_{p^2=m^2}=0,\quad\quad\quad a=\sigma,\pi.
\end{split}
\end{equation}
In the above equations, the coupling constants $K_\sigma=K_\pi=G_S+G_D\sigma_0/(16G_S)$ and the 1-PI loop functions $\Pi_a(p^2)=2\Pi_p(p^2), \Pi_\sigma(p^2)=2\Pi_s(p^2)$, with
\begin{equation}
\begin{split}
    \Pi_p(p^2)=\frac{i{\rm tr}S}{M(\sigma_0)}+\frac{1}{2}p^2I(p^2),\quad\quad\Pi^{ij}_s(p^2)=\frac{i{\rm tr}S}{M(\sigma_0)}+\frac{1}{2}[p^2-2M^2(\sigma_0)]I(p^2).
\end{split}
\end{equation}
At one loop, the loop functions $S$ and $I$ are given by
\begin{equation}
\begin{split}
    &i{\rm tr}S=\frac{N M(\sigma_0)}{2\pi^2}\bigg[\Lambda\sqrt{\Lambda^2+M(\sigma_0)}-M(\sigma_0)^2\log\frac{\Lambda+\sqrt{\Lambda^2+M^2(\sigma_0)}}{M(\sigma_0)}\bigg],
    \\
    &I(p^2)=\frac{2N}{\pi^2}\int_0^\Lambda \frac{dk k^2}{\sqrt{k^2+M^2(\sigma_0)}[4k^2+4M^2(\sigma_0)-p^2]}.
\end{split}
\end{equation}
While for the mesons $a=\eta, \eta'$, their masses are solutions of 
\begin{equation}
    \det[1-2\hat{K}_{\eta}\hat{\Pi}_{\eta}(p^2)]_{p^2=m^2}=0. 
\end{equation}
In the chiral limit, the matrices $\hat{K}_{\eta}$ and $\hat{\Pi}_{\eta}$ are  diagonal, 
\begin{equation}
\begin{split}
    \hat{K}_\eta=\Bigg(\begin{matrix}
    G_S-\frac{G_D\sigma}{4G_S}&0&\\
    0&G_s+\frac{G_D\sigma}{8G_S}&
    \end{matrix}\Bigg),
    \quad\quad
    \hat{\Pi}_\eta(p^2)=\Bigg(\begin{matrix}
    2\Pi_p(p^2)&0&\\
    0&2\Pi_p(p^2)&
    \end{matrix}\Bigg).
\end{split}
\end{equation}
As expected, among the three Goldstones, the $\pi$ and $\eta$ mesons are massless in the exactly chiral limit, and the mass of $\sigma$ is $2M(\sigma_0)$ if the anomaly term $G_D$ is turned off. In addition, the decay constant of $\pi$ can be computed by $f_\pi=M(\sigma_0)I(0)(\frac{\partial \Pi_\pi(0)}{\partial p^2})^{-1}$~\cite{Klevansky:1992qe}. The mesons for our reference points are shown in Table~\ref{table-1}. 


\begin{table} 
\begin{tabular}
{ |p{2.5cm}||p{1.8cm}|p{1.8cm}|p{1.8cm}|p{1.8cm}|p{1.8cm}|p{2cm}|   } 
 \hline
 \multicolumn{7}{|c|}{Parameter for $SU(3)$ NJL model} \\
 \hline
Benchmark& $f_\pi[\Lambda]$ & $m_\pi[\Lambda]$ &$m_\sigma[\Lambda]$&$m_{\eta}[\Lambda]$&$G_S[\Lambda^{-2}$]&$G_D[\Lambda^{-5}$]\\
 \hline
A& 0.384&0&0.769&0&2.0&0\\
B& 0.517&0&1.035&0&2.2&0\\
C& 0.637&0&1.274&0&2.4&0\\
D& 0.957&0&1.913&0&3.0&0\\
 \hline
 \multicolumn{7}{|c|}{Parameter for Polyakov Loop Model} \\
 \hline
Color number& $a_0$ & $a_1$ &$a_2$&$a_3$&$b_3$&$b_4$\\
 \hline
$N=3$& 6.75 & -1.95&  2.625  &-7.44    &0.75&7.5  \\
\hline
\end{tabular}
\caption{}  
\label{table-1}
\end{table}

\section{The Tunneling Potential Method for Chiral Phase Transition}\label{TunnelingPM}

The tunneling potential method~\cite{Espinosa:2018hue, Espinosa:2018szu} provides us with another approach to compute the bounce action for the tunneling equation modified by the non-trivial wave function of the order parameter field. It has the merit of estimating the bounce action quickly with a relatively good quality.

To establish the tunneling equation for this modified bounce equation \eqref{bounce}, notice that the bounce action \eqref{3daction} is split into two parts, the kinetic and potential parts, $S_E=S_K+S_V$, satisfying the relation $S_K=2S_E,\  S_V=-S_E$. The tunneling potential method finds a way to express $S_E$ in terms of the potential $V$  ($V_{PNJL}$ in this work) and the tunneling potential, which is defined in terms of the putative bounce solution to \eqref{bounce}, $\sigma_b$, as the following
\begin{equation}
    V_t=V-\frac{1}{2Z_\sigma}\dot{\sigma}_b^2, ~~{\rm with} ~~\dot{\sigma}_b=d\sigma_b/dr.
\end{equation}
The first and second derivative of $\sigma_b$ can be expressed in terms of $V$ and $V_t$
\begin{align}\label{s-boun}
    \dot{\sigma}_b&=-\sqrt{2Z_\sigma(V-V_t)},
   \\
   \ddot{\sigma}_b&=\frac{d\sigma_b}{dr}\frac{d}{d\sigma_b}\left(-\sqrt{2Z_\sigma(V-V_t)}\right)=Z_\sigma'(V-V_t)+Z_\sigma(V'-V_t').
\end{align}
Plugging them into \eqref{bounce}, we can express $r$ as a function of $V$ and $V_t$ 
\begin{equation}\label{r-boun}
    r=\frac{2\sqrt{2Z_\sigma(V-V_t)}}{-Z_\sigma V_t'}.
\end{equation}
Now, using Eq.~(\ref{s-boun}) and Eq.~(\ref{r-boun}), the bounce action will be known as long as $V_t$ is obtained:
\begin{equation}\label{VTaction}
    S_E=S_K/2=4\pi\int^{\sigma_+}_{\sigma_{e}}d\sigma \frac{[2Z_\sigma(V-V_t)]^{\frac{3}{2}}}{Z_\sigma^3V_t'^2},
\end{equation}
where the false vacuum $\sigma_+=0$  and the escaping point $\sigma_{e}$ will be found later. 

Then, the main task is to find $V_t$. First let us differentiate \eqref{r-boun} with $r$ on both sides, obtaining the bounce equation for $V_t$ 
\begin{equation}\label{boun}
    3Z_\sigma^2 V_t'^2=2Z_\sigma^2 V'V_t'-2(Z_\sigma Z_\sigma'V_t'+2Z_\sigma^2V_t'')(V-V_t).
\end{equation}
Recalling the original boundary condition $\dot\sigma_b|=0$ at $\sigma_{e}$ and $\sigma_b|_{r\rightarrow\infty}=0 \Rightarrow \frac{d\sigma_b}{dr}|_{r\to \infty }=0$ in \eqref{bounce} and using the definition \eqref{s-boun}, one can derive the two boundary conditions for $V_t$ 
\begin{equation}\label{b1}
    V_t(0)=V(0),\ \ \ \ V_t(\sigma_{e})=V(\sigma_{e}),
\end{equation}
which are irrelevant to derivative thus conducive to the numerical solution. Instead of solving \eqref{boun} with the boundary condition \eqref{b1} with brute force, one may try to construct the approximate tunneling potential $V_t$~\cite{Espinosa:2018hue}.

The key of the approach is to construct the trial $V_t$ that can reproduce the information on the boundaries and the top of the barrier. In addition to \eqref{b1}, we have the first derivatives on the boundaries: 1) $V_t'(0)=0$, obtained after plugging the extreme condition $V'(0)=0$ and the first of \eqref{b1} into \eqref{boun}; 2) $V_t'(\sigma_{e})=\frac{2}{3}V'(\sigma_{e})$ obtained after plugging the second of \eqref{b1} into \eqref{boun}. After the tedious algebraic calculations, one obtains the trial potential 
\begin{equation}
\begin{split}
V_{t4}=V_{t}^{(1)}+V_{t}^{(2)}+V_{t}^{(3)}+V_{t}^{(4)},
\end{split}
\end{equation}
with
\begin{equation}
\begin{split}
    &V_{t}^{(1)}=V_0\frac{\sigma}{\sigma_{e}},
    \\
    &V_{t}^{(2)}=\left(\frac{2}{3}V_0'\sigma_{e}-V_0\right)\frac{\sigma(\sigma-\sigma_{e})}{\sigma_{e}^2},
    \\
    &V_t^{(3)}=\bigg(\frac{2}{3}V_0'\sigma_{e}-2V_0\bigg)\frac{\sigma(\sigma-\sigma_{e})^2}{\sigma_{e}^3},
    \\
    &V_t^{(4)}=a_4\sigma^2(\sigma-\sigma_{e})^2,
\end{split}
\end{equation}
where we have assumed $V(\sigma_+)=0$ and $V(\sigma_{e})=V_0$. The unknown parameter $a_4$ can be eliminated by considering \eqref{boun} at the top of the barrier which is located at $\sigma_T$,
\begin{equation}
3Z_\sigma(\sigma_T)V_{t4}'(\sigma_T)^2=-2[Z_\sigma(\sigma_T)Z_\sigma'(\sigma_T)V_{t4}'(\sigma_T)+2Z_\sigma^2(\sigma_T)V_{t4}''(\sigma_T)][V(\sigma_T)-V_{t4}(\sigma_T)].
\end{equation}
Now, substituting $V_{t4}$ into \eqref{VTaction}, one gets $S_E(\sigma_{e})$. Finally, searching for the minimum of $S_E(\sigma_{e})$ determines the escaping point thus the tunneling potential. 
 

\bibliographystyle{unsrt}  
\bibliography{main}

\end{document}